\documentclass[aps]{revtex4-1}
\usepackage{graphicx,color}
\usepackage{amsmath,amssymb}
\usepackage{bm,theorem}

\newtheorem{theorem}{Theorem}
\newcommand{\vv}{\mathbf{v}}
\newcommand{\newc}{\newcommand}
\newc{\be}{\begin{equation}}
\newc{\ee}{\end{equation}}
\newc{\bea}{\begin{eqnarray}}
\newc{\eea}{\end{eqnarray}}
\newc{\beas}{\begin{eqnarray*}}
\newc{\eeas}{\end{eqnarray*}}
\newc{\pard}{\partial}
\newc{\ti}{\tilde}
\newc{\tr}{\textrm}
\newc{\ds}{\displaystyle}
\newc{\noi}{\noindent}

\begin{document}

\title{Classical and quantum vortex leapfrogging in two-dimensional channels}

\author{Luca Galantucci}
\author{Michele Sciacca}
\author{Nick G. Parker}
\author{Andrew W. Baggaley}
\author{Carlo F. Barenghi}

\affiliation{Joint Quantum Centre Durham--Newcastle, School of Mathematics,
Statistics and Physics, Newcastle University, Newcastle upon Tyne, NE1 7RU, United Kingdom}


\begin{abstract}
The leapfrogging of coaxial vortex rings is a famous effect which has
been noticed since the times of Helmholtz. Recent advances in 
ultra-cold atomic gases show that the effect can now be studied in 
quantum fluids. The strong confinement which characterizes these systems
motivates the study of leapfrogging of vortices within narrow channels.
Using the two-dimensional point vortex model, we show that in the constrained
geometry of a two-dimensional channel the dynamics is richer than in an 
unbounded domain: alongsize the known regimes of standard leapfrogging and the
absence of it, we identify new regimes of
backward leapfrogging and periodic orbits. 
Moreover, by solving the Gross-Pitaevskii equation for a Bose-Einstein
condensate, we show that all four regimes exist for quantum vortices too.
Finally, we discuss the differences between classical and quantum vortex
leapfrogging which appear when the quantum healing length becomes
significant compared to the vortex separation or the channel
size, and when, due to high velocity, compressibility effects in the
condensate becomes significant.
\end{abstract}

\maketitle


\section{Introduction}

The leapfrogging of two co-axial vortex rings (in three dimensions) or
of two vortex-antivortex pairs (in two dimensions)
is a benchmark problem of vortex interaction 
\citep{Meleshko2010} which dates back to \citep{Helmholtz1858}. 
The time evolution of this vortex
configuration is striking: 
the vortex ring (or pair) which is ahead widens and slows
down, while the ring behind contracts,
speeds up, catches
up with the first ring and goes ahead through it; 
this `leapfrogging' game is then repeated over and over again, 
unless instabilities disrupt it. 
A number of papers have been written on different
aspects of this problem,
ranging from the stability 
\citep{Hicks1922,Love1894,Acheson2000,TophojAref2013}  
to the deformation of the vortex
cores and to the effects of viscosity \citep{ShariffLeonard1992}
using numerical 
\citep{RileyStevens1993,Borisov2014,ChengLim2015}
as well as experimental methods 
\citep{Maxworthy1972,YamadaMatsui1978,Lim1997,Qin2018}. The
most recent developments concern leapfrogging of
vortex bundles \citep{Wacks2014} and
helical waves \citep{Hietala2016,Selcuk2018,Quaranta2019}.

Our work is motivated by recent experiments with atomic Bose-Einstein 
condensates, which constitute a dilute quantum fluid and provide an idealised platform to study fundamental 
vortex dynamics \citep{White2014}.
In these experiments,
atomic gases are confined by suitable magnetic-optical traps and cooled 
to nano-Kelvin temperatures.
If the atoms of the gas are bosons (i.e. have integer spin), 
a phase transition occurs upon cooling below 
a critical temperature $T_c$,
and the gas forms a macroscopic coherent quantum state 
\citep{BarenghiParker2016} called a Bose-Einstein condensate (BEC). 
From the point of view of the hydrodynamics, a BEC has
three key properties: it is superfluid (i.e. it suffers no  
viscous losses of kinetic energy when it flows), it is compressible, and
its vorticity is concentrated to thin hollow vortex lines with fixed
width $a_0$ and fixed  
circulation $\pm h/m$ where $h$ is Planck's
constant and $m$ is the mass of a boson (while vortices with larger quanta of circulation, 
$\pm 2 h/m, \pm 3 h/m, \cdots$, are possible, they are unstable to decay into multiple singly-charged vortices).  
Thus, in BECs, vortices are well-defined and identical objects, 
evolving in an inviscid compressible fluid.


There are several additional characteristics of atomic BECs that make them attractive for probing vortex dynamics.  Firstly, the physical parameters of the fluid (including the width and speed of the vortices) are tunable, for example, through the density of the gas and the strength of the atom-atom interaction (which can be modified by means of Feshbach resonances \citep{Inouye1998}); this should be contrasted with superfluid liquid helium - historically the most studied quantum fluid - whose physical parameters are fixed by nature.  Secondly, the potential experienced by the gas can be controlled through magnetic and optical fields.  Such trapping is essential, on one hand, to contain the gas, and gives rise to the boundary effects which are central to this work.  However, the potential can also be exploited to engineer the dimensionality of the gas - particularly, quasi-two-dimensional geometries in which vortex lines effectively become point-like vortices - and to stir and shake the condensate.  Finally, recent techniques have enabled the observation of vortex lines \citep{Serafini2017} and vortex points \citep{Seo2017}  in real-time, including inference of their individual circulations.  
%


Atomic BECs have been employed as a context to study a range of fundamental vortex phenomena, 
including vortex nucleation from moving obstacles 
\citep{frisch-etal-1992,nore-etal-1993,Neely2010,Stagg2014,Kwon2015} 
and flow constriction \citep{Valtolina2015,Burchianti2018,xhani-etal-2020}, 
von K\'arm\'an vortex 
streets \citep{Sasaki2010,Kwon2016}, 
vortex-antivortex annihilations \citep{Seo2017}, 
vortex line reconnections \citep{Serafini2017,Galantucci2019}, 
vortex chaos \citep{Navarro2013}, vortex scattering 
\citep{Barenghi2005,Caplan2014,Griffin2017}, 
quantum turbulence \citep{Henn2009,Neely2013,Kwon2014,Stagg2015,White2014,Tsatsos2016,garciaorozco-etal-2020}, 
and self-organisation and clustering of vortices \citep{Simula2014,Billam2014,Gauthier2019,Johnstone2019}.
With regards to vortex leapfrogging, this has been considered theoretically in idealised unconfined condensates \citep{Ikuta2019}, including
spinor condensates \citep{Kaneda2014}.


Atomic BECs however are characterized by their small dimensions, 
typically from 10 to 100 times the vortex core size, for which the
motion of vortices can be significantly affected by the presence of
boundaries. This drawback is mainly due to the loss of atoms in 
the final evaporative stage of cooling the gas. 
There are even experiments
in which, by design, the most interesting physics occurs in the most
restricted region of the system, for example vortex rings nucleated
in the weak link of the Josephson junction between two condensates
\citep{Valtolina2015,xhani-etal-2020}. 
The aim of the present work is to provide insight in the interpretation 
of current and future experimental 
studies of vortex dynamics in 
confined condensates (rather than idealised open domains), 
where leapfrogging dynamics, which can be established if the 
vortex nucleation frequency is sufficiently high, is affected by the
presence of boundaries. 
The characteristics of leapfrogging motion in such confined systems
is likely to show significant dissimilarities compared to the corresponding dynamics in unbounded systems stemming from 
the role played by image vortices arising from the presence of boundaries. Despite the expected impact of geometrical 
confinement, to the best of our knowledge the role of boundaries in leapfrogging dynamics has never been investigated in 
literature neither for classical nor for quantum fluids (\citep{Kaneda2014} and
\citep{Ikuta2019} indeed studied leapfrogging in homogeneous condensates, without boundaries).
In order to assess the impact of
the boundaries and disentangle the latter from other concurrent physical 
effects existing in quantum fluids (\textit{e.g.} compressibility), 
in this research we compare the leapfrogging of vortices in plane channels in (i) ideal incompressible classical fluids 
and (ii) box-trapped Bose-Einstein condensates. 
In order to simplify the system under investigation, our theoretical and numerical analysis is performed in
two-dimensions, employing the point vortex model for classical fluids and 
the Gross-Pitaevskii equation for BECs. We stress that 
the Gross-Pitaevskii equation
has proved an excellent quantitative model of experiments
with Bose-Einstein
condensates at temperatures $T \ll T_c$; at relatively high values of
temperature, the condensate exchanges energy and particles with the
thermal cloud, and the Gross-Pitaevskii equation requires modifications 
\citep{Brewczyk2007,Proukakis2008,Blakie2008,Berloff2014}.
We also remark that on one hand the two-dimensional nature of the
system that we consider is an idealisation (the aim is to get insight
in the motion of three-dimensional vortex rings), but, on the other
hand, where atomic Bose-Einstein condensates are
tightly confined in one direction the system becomes effectively 
two-dimensional and our  two-dimensional approach becomes realistic.


The article is organised as follows. In Section \ref{sec:models}, 
we illustrate the two theoretical models employed,
namely the classical point vortex model and the Gross-Pitaevskii equation 
describing the dynamics of BECs in the zero-temperature
limit. In Section \ref{sec:results}, we report the results obtained 
in both classical and quantum fluids, focusing on the 
role of boundaries and on the differences between classical and 
quantum systems. 
Finally, in the last Section \ref{sec:concl}, we summarise our
findings and illustrate their importance in the future of quantum 
vortex experiments.

\section{Models}
\label{sec:models}

\subsection{Point vortex model}
 
The simplest model of our system is the classical point vortex model:
a two-dimensional inviscid incompressible irrotational
fluid in an infinite channel of width $2D$ containing two 
vortex-antivortex pairs
(the two-dimensional analog of three-dimensional coaxial vortex rings),
each of circulation $\pm\Gamma$. 
In view of comparing the results obtained with this classical model 
to quantum vortices in confined BECs,
the hypotheses behind the point vortex model must be carefully considered. 

The classical model describes a fluid with constant density. 
In the bulk of the condensate, 
{\it i.e.} sufficiently far from boundaries or vortices, 
this assumption is 
realistic: indeed, although in past experiments condensates were 
usually confined by harmonic trapping potentials resulting 
in density gradients \citep{dalfovo-etal-1999}, 
current experimental techniques \citep{Gaunt2013} allow 
box-like trapping potentials which lead to uniform density 
profiles in the bulk of the condensate as in the classical
point vortex model.
In particular, in the vicinity of a vortex,
the classical model assumes constant 
density at any radial distance $r$ to the vortex axis, including the
vortex axis $r=0$ itself. 
In Bose-Einstein condensates, a vortex is a topological defect
of the phase of the governing complex wavefunction (or order parameter),
as we shall describe with more details in Section~\ref{sec:theory.quantum}.
Therefore the vortex core is a thin tubular region around the vortex
axis which is depleted of atoms: as $r \rightarrow 0$, the velocity
tends to infinity, as in the point vortex model, but the fluid density 
tends to zero.  The radius of this tube is of the 
order of the quantum mechanical healing length $\xi$ 
(see Section \ref{sec:theory.quantum}). A similar difference between the
classical point vortex model and Bose-Einstein condensates occurs
near a hard boundary: the classical model assumes that the fluid's density is
constant up to the boundary; in a Bose-Einstein condensate
a thin boundary region (again of the order of $\xi$) forms near the
boundary where, in the case of box-like traps, the condensate's density 
rapidly drops from the bulk value to zero. 
We conclude that, from a geometrical point of view,
the classical point vortex model can be used to model
Bose-Einstein condensates provided that vortex-vortex and vortex-boundary
distances are larger than the healing length $\xi$.

From a dynamical point of view,
the assumption of constant density implies that
the classical point vortex model 
neglects sound waves which are radiated away by quantum vortices 
when they accelerate \citep{Barenghi2005}. 
The point vortex model, in fact, is based on the classical ideal Euler
equation which conserves energy. In the low temperature limit $T/T_c \ll 1$
that guarantees the validity of the Gross-Pitaevskii equation, the
total energy of a BEC is constant, but transformation of incompressible 
kinetic energy of the vortex configuration into compressible kinetic
energy of the field of sound waves (or vice versa) is permitted.
This dynamical difference between the classical point vortex model and
the Gross-Pitaevskii model is, physically, perhaps the most significant,
and will be addressed while discussing the results
in Section \ref{sec:results}.


Despite these approximations, we reckon the model captures
the essential ingredient of our problem: the motion of quantised irrotational vortices
in presence of boundaries.
Indeed, the classical point vortex model in a circular disk has been 
already used with success to model two-dimensional
turbulence in low temperature 
trapped condensates, for example, by \citet{Simula2014}. 
It must also be noticed that \citet{Mason2006}
have shown that the motion of a realistic vortex at distance $d$ to
a boundary can be described in terms of a classical image vortex even 
if $\xi$ is comparable to $d$ (although a small correction is needed to
account for the density depletion in the boundary region).
In the suitable physical limits, we hence expect the point vortex model to correctly 
describe the impact of boundaries on the leapfrogging
of quantised vortices.

\subsubsection{Equations of motion}

Our physical domain under investigation is a two-dimensional infinite strip $\mathcal{C}_{\mathbb{R}^2}$ defined as 
$\mathcal{C}_{\mathbb{R}^2} =\{ (x,y) \in \mathbb{R}^2 \;\; : \;\; (x,y) \in \left ( -\infty , \infty \right ) \times  \left ( 0 , 2 D \right ) \} $, 
which hereafter we will refer to as the {\it channel}. 
We assume the flow to be two-dimensional, {\it i.e.} the velocity vertical component $v_z=0$ and the horizontal components
$v_x$ and $v_y$ only depend on horizontal coordinates $x$ and $y$ and time $t$. The incompressibility assumption implies that the 
continuity equation can be written as follows 
\begin{equation}
\displaystyle
\nabla \cdot \mathbf{v} = 0. \label{eq:div_free}
\end{equation}
The velocity field $\vv$ can hence be expressed as the curl of vector field $\displaystyle\boldsymbol{\Psi}$ which, 
given the two-dimensionality of the flow,
has non-vanishing components only in the $z$ direction, $\displaystyle\bm{\Psi} = (0,0,\psi(x,y,t))$. The velocity components have hence
the following expressions in terms of the function 
$\psi$ which is often denominated {\it streamfunction}: $v_x~=~\partial_y \psi$ and $v_y~=~-\partial_x \psi$, where
$\partial_i$ indicates spatial derivatives in the $i$ direction. 

The irrotationality of the flow implies that the velocity field can be expressed via a {\it potential} function $\varphi$, {\it i.e.}
\begin{equation}
\displaystyle
\mathbf{v} = \nabla \varphi \;\; ,\label{eq:curl_free}
\end{equation}
leading to the following relations for the components $v_i=\partial_i \varphi$. Equations~(\ref{eq:div_free}) and (\ref{eq:curl_free}) 
imply 
that both $\varphi$ and $\psi$ satisfy Laplace equation, $\Delta \varphi = \Delta \psi = 0$, and the 
following equalities between their spatial derivatives:
\begin{eqnarray}
\displaystyle
\partial_x \varphi & = & \partial_y \psi \;\; ,\label{eq:cauchy_riemann_1}\\
\partial_y \varphi & = & -\partial_x \psi. \label{eq:cauchy_riemann_2}
\end{eqnarray}
Equations~(\ref{eq:cauchy_riemann_1}) and (\ref{eq:cauchy_riemann_2}) coincide with the well known Cauchy-Riemann relations for
the complex function $\Omega(z) := \varphi + \rm i \psi$, where $z=x + i y$. Hence, following basic complex analysis, 
the function $\Omega(z)$, denominated 
{\it complex potential}, is an analytical complex function on the simply connected open domain 
$\mathcal{C}=\{ z\;\in\;\mathbb{C}\; :\;  0 < \Im\rm m \; z < 2D \} \subsetneq \mathbb{C}$.
As a consequence, $\Omega(z)$ is differentiable and its derivative 
\begin{equation}
\displaystyle
w(z) := \frac{d\Omega}{dz} = v_x - i v_y \label{eq:def_cplx_vel}
\end{equation} 
is the so-called 
{\it complex velocity}. In the framework of complex potentials, the impermeable boundary conditions for ideal fluids
correspond in our channel $\mathcal{C}$ to the following constraint: 
$ \Im\text{m}\; \Omega(z)_{\arrowvert_{z\in\partial\mathcal{C}}} = \alpha(t)$, with $\alpha(t)\in\mathbb{R}$ depending only on time 
$t$.

The description of incompressible and irrotational flows of ideal fluids via the complex potential-based formulation is particularly
useful in the present work as it allows the employment of {\it conformal mapping techniques} for the derivation of the analytical
expression of the complex potential $\Omega(z)$ describing the velocity field induced by a point vortex in our channel $\mathcal{C}$. 
The essential steps for this derivation are as follows. The necessary ingredients are mainly two: (a) the knowledge of the complex potential
$\Theta(\zeta)$ describing the flow induced by a point vortex in a simply connected open subset $\mathcal{D}$ of the complex plane,
with $\zeta\in\mathcal{D}\subsetneq \mathbb{C}$; and (b) the construction of a conformal map $\zeta=f(z)$ transforming our channel $\mathcal{C}$ onto
the domain $\mathcal{D}$. 

Conformal maps $f$ are transformations defined on the complex plane which preserve angles. 
Such maps are performed by analytical complex functions with non-vanishing derivative, 
{\it i.e.}, in the present case, $f'(z)\neq 0$ for all $z\in \mathcal{C}$. The requirement $\mathcal{D}$ not
coinciding with the entire complex plane $\mathbb{C}$, is fundamental in order to exploit the Riemann Mapping Theorem which ensures the 
existence of the conformal map $f$ mapping $\mathcal{C}$ onto $\mathcal{D}$. 
Once $\Theta(\zeta)$ and $f(z)$ are determined, the complex potential $\Omega(z)$ for a vortex flow in $\mathcal{C}$
is obtained by transforming the potential $\Theta(\zeta)$ via the conformal map $f^{-1}(\zeta)$, {\it i.e.}
\begin{equation}
\displaystyle
\Omega(z)=\Theta(f(z)) \;\; . \label{eq:omega_general}
\end{equation}
The reasons why the derived complex function $\Omega(z)$ via Eq. (\ref{eq:omega_general}) is the seeked complex potential are the following.
First, $\Omega(z)$ is analytic on $\mathcal{C}$
(as it is obtained via the composition of two analytic functions, $f$ and $\Theta$), implying that the real and imaginary
parts of $\Omega(z)$ are related to each other via the Cauchy-Riemann equations and are both harmonic functions. Hence, they 
do satisfy all the necessary conditions for corresponding respectively to a potential function and a streamfunction
of an incompressible and irrotational flow of an inviscid fluid.
Second, the correspondence of $\partial \mathcal{C}$
and $\partial \mathcal{D}$ under the conformal mapping performed by $f$ transposes the boundary conditions enforced by $\Theta(\zeta)$ on 
$\partial \mathcal{D}$ to the boundary $\partial \mathcal{C}$ \citep{lavrentiev-chabat-1972}. 
Finally, via conformal mappings, the flow induced by a vortex of circulation $\kappa$
is indeed mapped to a vortex flow with the same circulation \citep{newton-2001}.

In the present work, we choose $\mathcal{D}$ to coincide 
with the upper half complex plane, {\it i.e.} $\mathcal{D}=\{ \zeta\in\mathbb{C} : \Im\text{m} \; \zeta > 0 \}$. 
In this domain, the complex potential $\Theta(\zeta)$
describing the flow induced by a vortex placed in $\zeta_0\in\mathcal{D}$ is obtained by the method of images, namely
\begin{equation}
\displaystyle
\Theta(\zeta,\zeta_0) = -\text{sgn}(\zeta_0)\frac{i\kappa }{2 \pi}
\log{\left( 
\frac{\zeta -\zeta_0}{\zeta - \zeta_0^*}\right)} \;\; ,\label{eq:cplx_pot_upper_half}
\end{equation}
where $\text{sgn}(\zeta_0)$ is the sign of the vortex placed in $\zeta_0$ (positive for anti-clockwise induced flow, 
negative for clockwise), $\zeta_0^*$ is the complex conjugate of $\zeta_0$ where a vortex of opposite sign
is placed (the image-vortex of $\zeta_0$) and $\kappa$ is the circulation of the flow generated by the vortex. 
\begin{figure}
\begin{center}
\includegraphics[width=0.95\textwidth]{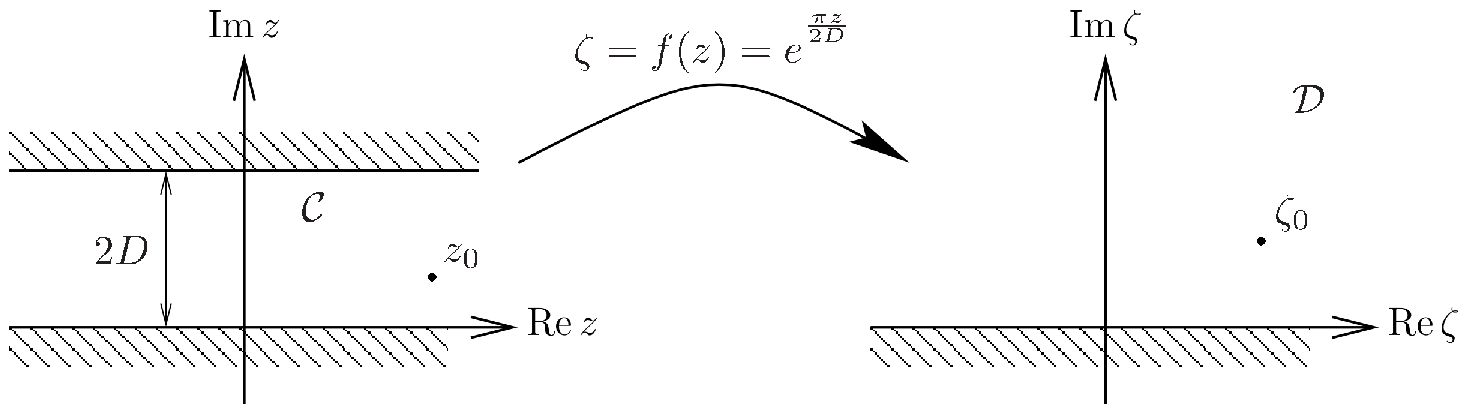}
\caption{Schematic illustration of the conformal map $\zeta=f(z)= e^{^{\frac{\pi z}{2D}}}$ transforming 
$\mathcal{C}$ onto $\mathcal{D}$ and a vortex placed in $z_0$ into a vortex in $\zeta_0$, $\zeta_0=f(z_0)$.}
\label{fig:schematic_conf_maps}
\end{center}
\end{figure}
The analytical function $f$ transforming conformally the channel  
$\mathcal{C}=\{ z\in\mathbb{C} : 0 < \Im\text{m} \; z < 2D \}$ onto $\mathcal{D}$ is as follows (see Fig.~\ref{fig:schematic_conf_maps}
for a schematic illustration)
\begin{equation}
\displaystyle
\zeta=f(z)= e^{^{\frac{\pi z}{2D}}} \;\; . \label{eq:conformal_map_f}
\end{equation}
The conformal map $f$ transforms $\partial \mathcal{C}$ onto $\partial \mathcal{D}$, with $f(\{ z\in\mathbb{C} : \Im\text{m} \;z= 0\})=\mathbb{R}^+$ and
$f(\{ z\in\mathbb{C} : \Im\text{m} \; z= 2D\})=\mathbb{R}^-$. Employing Eq. (\ref{eq:omega_general}), the determination of the 
complex potential $\Omega(z)$ is straightforward, namely 
\begin{equation}
\displaystyle
\Omega(z,z_0)=-\text{sgn}(z_0)\frac{i\kappa}{2\pi}
\log{\left(\frac{1-e^{^{-\frac{\pi}{2D}(z-z_{0})}}}{1-e^{^{-\frac{\pi}{2D}(z-z_{0}^*)}}}\right)} \;\;\; ,
\;\;\; z_0=f^{-1}(\zeta_0)
\label{eq:cplx_omega}
\end{equation}
leading to the following complex velocity 
\begin{eqnarray}
\displaystyle
w(z,z_0)&=&-\text{sgn}(z_0)\frac{i\pi}{4}\frac{\kappa}{2\pi D}\left \{ \coth \left [ \frac{\pi}{4D}(z-z_0)\right ]-
\coth \left [ \frac{\pi}{4D}(z-z_0^*) \right ] \right \} \nonumber \\[2mm]
&=& \chi(z,z_0) + \chi(z,z_0^*) \;\; , \label{eq:cplx_vel}
\end{eqnarray}
where  
$\displaystyle\;\chi(z,z_0)=-\text{sgn}(z_0)\frac{i\pi}{4}\frac{\kappa}{2\pi D} \coth \left [ \frac{\pi}{4D}(z-z_0)\right ]$ 
and $\;\text{sgn}(z_0^*)=-\text{sgn}(z_0)$. \\[2mm]
The complex function $\chi(z,z_0)$ (and, correspondingly, $\chi(z,z_0^*)$) 
can be physically interpreted as the complex velocity generated by an isolated vortex placed in $z_0$ 
(whose complex potential would be $\Omega(z,z_0)=-\text{sgn}(z_0)i\kappa\log(z-z_0)/(2\pi)$)
and its infinite images with respect to the walls of the channel, $\Im\text{m} \; z=0$ and $\Im\text{m} \; z=2D$.
The expression (\ref{eq:cplx_vel}) for the complex potential $w(z,z_0)$ can indeed be derived by considering two sets of 
infinite images of a vortex placed in $z_0$ and an anti-vortex in $z_0^*$ \citep{greengard-1990}.

If the channel is characterised by the presence of $N$ vortices, the complex velocity $w(z,z_{k_{\{k=1,\dots,N\}}})$ generated by
the the set of $N$ vortices is obtained via the superposition principle, {\it i.e.}
\begin{equation}
\displaystyle
w(z,z_{k_{\{k=1,\dots,N\}}})=\sum_{k=1}^N w(z,z_k)=\sum_{k=1}^N\left [ \chi(z,z_k) + \chi(z,z_k^*) \right ] \;\; .
\label{eq:cplx_vel_N}
\end{equation}
A crucial role in this $N$-vortex problem is played by the equations of motion of a generic $j$-th vortex. In order to derive
such equations of motions, we define the position $z_j(t):=x_j(t) + i y_j(t)$ occupied by the vortex at time 
$t$ in the channel $\mathcal{C}$. Indicating with the superscript ` $\dot{}$ ' derivation with respect to time, we define the 
quantity $\dot{z}_j(t):=\dot{x}_j(t) + i \dot{y}_j(t)$, where the real and imaginary part correspond to the 
$x$ and $y$ components of the $j$-th vortex velocity. As vortices are advected by the local fluid velocity, {\it i.e.} 
$\dot{\mathbf{x}_j}(t)= \mathbf{v}(\mathbf{x}_j(t),t)$, the following relation holds 
%
\begin{equation}
\displaystyle
\dot{z}_j=w^*(z_j,z_{k_{\{k=1,\dots,N\}}}) \;\; ,
\label{eq:cplx_vel_j}
\end{equation}
where we have omitted the time dependence of $z_j$ and $z_k$ to ease notation and the complex conjugation on the r.h.s.
arises from the definition (\ref{eq:def_cplx_vel})  of complex velocity. In order to determine the complex velocity 
$w(z_j,z_{k_{\{k=1,\dots,N\}}})$, we employ Eq. (\ref{eq:cplx_vel_N}) subtracting the term corresponding to the vortex placed in 
$z_j$, obtaining the following relation
\begin{eqnarray}
\displaystyle
\dot{z}_j & = & w^*(z_j,z_{k_{\{k=1,\dots,N; k\neq j\}}}) + \chi^*(z_j,z_j^*) \nonumber \\[2mm]
          & = & \sum_{k\neq j} w^*(z_j,z_k) + \chi^*(z_j,z_j^*) \nonumber \\ 
          & = & \sum_{k\neq j} \left [ \chi^*(z_j,z_k) + \chi^*(z_j,z_k^*) \right ] + \chi^*(z_j,z_j^*) \;\; ,
\label{eq:motion_j}
\end{eqnarray}
which coincides with the equations of motion of the $j$-th vortex. The equations of motion of the whole $N$-vortex problem
are hence a set of $2N$ coupled ordinary differential equations.

\subsection{Gross-Pitaevskii equation model}
The Gross-Pitaevskii model is a well-established theoretical framework for the investigation of the
dynamics of BECs at temperatures much smaller
than the critical transition temperature. The Gross-Pitaevskii (GP) equation describes the 
temporal evolution of the complex order parameter 
$\Psi= \Psi(\mathbf{x},t)$ of the system, and reads as follows,

\begin{equation} 
\displaystyle
i\hbar \dot{\Psi} = -\frac{\hbar^2}{2m}\Delta \Psi + V \Psi + g |\Psi|^2 \Psi \;\; ,
\label{eq:GP}
\end{equation}

\noindent
where the dot is the time derivative,
$\hbar=h/(2\pi)$ is the reduced Planck's constant, $m$ is the boson mass, $V=V(\mathbf{x},t)$ is an externally  applied
potential, and $g=4\pi \hbar^2 a_s/m$ models the two-body contact-like boson interaction, where $a_s$ is the s-wave scattering length
for the collision of two bosons. 
The order parameter $\Psi$ can be written in terms of its amplitude
and its phase as

\begin{equation}
\displaystyle
\Psi = \sqrt{n} e^{i\theta}\; ,
\label{eq:psi_polar}
\end{equation}

\noindent
where $n=n(\mathbf{x},t)=|\Psi|^2$ is the particle number density 
(number of bosons per unit volume) and $\theta=\theta(\mathbf{x},t)$ is the
phase. 
Without loss of generality, the order parameter $\Psi$ can be written
as $\Psi(\mathbf{x},t)=e^{i \mu t/\hbar} \Phi(\mathbf{x},t)$ where
$\mu$ is called the chemical potential and $\Phi(\mathbf{x},t)$ obeys

\begin{equation} 
\displaystyle
i\hbar \dot{\Phi} = -\frac{\hbar^2}{2m}\Delta \Phi + V \Phi + g |\Phi|^2 \Phi
-\mu \Phi \;\; .
\label{eq:GPmu}
\end{equation}

\subsubsection{Quantum vortices}
\label{sec:theory.quantum}

In the context of BECs described by the 
Gross-Pitaevskii equation, quantum vortices are topological defects 
of the phase $\theta$ of the order parameter, at which
$\Psi=0$ (hence $\theta$ is undefined)
and around which $\theta$ wraps by $2q\pi$ 
with $q\in\mathbb{Z}\setminus \{ 0 \}$. 
In three dimensions, vortices take the form of one-dimensional curves which may form a vortex tangle, as observed both in BECs \citep{White2010} 
and superfluid helium \citep{Vinen1957}. In two dimensions, vortices coincide with vortex points which have been observed 
extensively in oblate (pancake-like) BECs \citep{matthews-etal-1999}. For the purpose of the present work, we will restrict our discussion to two dimensional systems. 

The velocity field $\mathbf{v}(\mathbf{x},t)$ associated to a BECs whose dynamics is described by the order parameter $\Psi$, is defined
from the phase $\theta$ via the relation
\begin{equation}
\displaystyle
\mathbf{v}(\mathbf{x},t)=\frac{\hbar}{m}\nabla \theta.
\label{eq:v_grad_phi}
\end{equation}
Employing the definition (\ref{eq:v_grad_phi}) of the velocity and the $2q\pi$ phase wrapping existing around a vortex,
it is straightforward to verify that the circulation $\Gamma$ of 
the velocity field on any closed curve $\gamma$ enclosing a vortex point is quantised in terms of the quantum of circulation $\kappa=h/m$, {\it i.e.}
\begin{equation}
\displaystyle
\Gamma = \oint_\gamma \!\! \mathbf{v}\cdot \mathbf{dl}=q \kappa \;\;\; , \;\;\; q\in\mathbb{Z}\setminus \{ 0 \}.
\label{eq:quant_circ}
\end{equation}
Choosing $\gamma$ to be a circle of radius $r$ and assuming the flow around a vortex to be axisymmetric,
the azimuthal component of the flow velocity around a vortex is given by the relation $v_\phi=q\kappa/(2\pi r)$,
coinciding with the expression for a classical point vortex. Hence, from a velocity point of view, quantum and 
classical vortices are identical. The important and dynamically significant distinction between classical and 
quantum vortices is that the latter are characterised by a finite core whose size is of the order of the so-called {\it healing length}
$\xi=\hbar/\sqrt{mgn}$. As we will very briefly illustrate in the next section, quantum fluids are indeed compressible fluids.

\subsubsection{Fluid dynamical equations for a BEC}
The Gross-Pitaevskii equation (\ref{eq:GP}) may be rewritten via the {\it Madelung transformation} consisting in 
expressing $\Psi$ in polar form (\ref{eq:psi_polar}) and separating the real and imaginary parts of (\ref{eq:GP}). 
This procedure leads to the following 
equations
\begin{eqnarray}
\displaystyle
&& \dot{n} + \nabla\cdot \left ( n\mathbf{v}\right ) = 0, \label{eq:quantum_continuity}\\[3mm]
&& m n \left [ \dot{\mathbf{v}} + \left ( \mathbf{v}\cdot\nabla \right ) \mathbf{v} \right ] =
-\nabla \left ( p + p'\right ) -n\nabla V, \label{eq:quantum_moment}
\end{eqnarray}
where $p$ and $p'$ are respectively {\it pressure} and {\it quantum pressure}
\begin{eqnarray}
\displaystyle
&&  p = \frac{g n^2}{2},\\[3mm]
&&  p'= -\frac{\hbar^2}{4m} n \Delta \left ( \ln{(n)} \right ).
\end{eqnarray}
Equation~(\ref{eq:quantum_continuity}) coincides formally with the continuity equation of a classical fluid, while 
equation (\ref{eq:quantum_moment}), exception made for the presence of the quantum pressure $p'$, 
is formally identical to the momentum balance equation for a barotropic, compressible classical Euler (ideal) fluid. At 
length scales $\ell$ much larger than the healing length $\xi$ (which is the typical length scale for density variations, associated {\it e.g.}
to the presence of vortices or boundaries) $p'/p \ll 1$, implying that in this limit the BEC can indeed be considered
as a barotropic, compressible classical inviscid fluid. Hence, at length scales $\ell\gg \xi$, the dynamics of quantum and classical 
point vortices only differ on the basis of compressible phenomena which may arise in BECs. In the other limit of $\ell\sim \xi$, the physics
may be significantly different. For instance, if the relative distance between quantum vortices of opposite sign is of the order of $\xi$,
the quantum pressure term would trigger the annihilation of the vortex pair, while in the classical point vortex model no loss of circulation is
included in the model. Moreover, the behaviour of a co-rotating pair of quantum vortices of same sign also shows dissimilarities with respect to the
classical case, in particular for the finite value of the rotation frequency $\omega_\tau$ 
as the distance $\ell$ tends to zero (in the classical model, the frequency diverges, $\omega_\tau \sim 1/\ell^2$).

\section{Results}
\label{sec:results}

\subsection{Classical fluids}
\label{sec:results.classical}

\begin{figure}
\begin{center}
\includegraphics[width=0.95\textwidth]{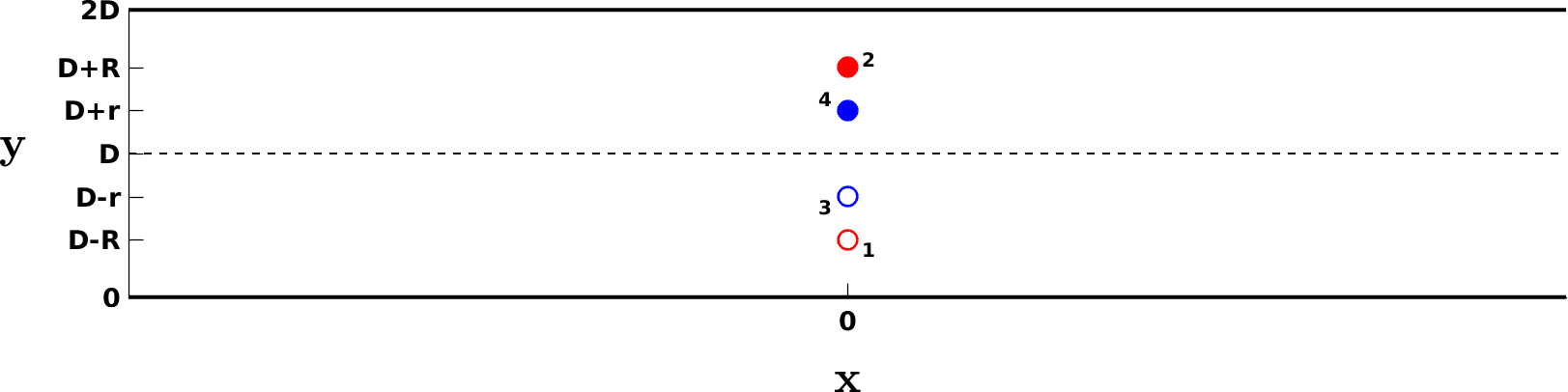}
\caption{Initial vortex configuration for the classical point vortices numerical simulations: filled (open) circles correspond to 
vortices with positive (negative) circulation. Numerical labels close to vortices indicate the vortex numeration employed.}
\label{fig:ic_classical}
\end{center}
\end{figure}

To make progress in understanding the impact of boundaries on the leapfrogging behaviour of classical point vortices in a two-dimensional channel, 
we consider the motion of four vortices,  half with positive circulation
$\kappa$, half with negative $-\kappa$.
In Fig.~\ref{fig:ic_classical} we show this
initial condition. If we interpret our two-dimensional configuration
as a model of a three-dimensional configuration of vortex rings,
point vortices of same colour in the figure correspond to cross-sections 
of the same ring.
%
%
Initially, the four vortices are vertically aligned on the $y$ axis, {\it i.e.} $x_j(0)=0$ for $j=1,\dots,4$ and the vortex-anti vortex 
pairs are symmetrically positioned with respect to the channel mid axis $y=D$, namely $y_j(0)=D \pm R$ for the first pair $j=(1,2)$ 
and $y_j(0)=D \pm r$ for the second pair $j=(3,4)$, with the conditions $R/D < 1 $ and $r/R < 1$.
In order to characterise the dependence of vortex trajectories on the two non-dimensional parameters $r/R$ and $R/D$ which determine the flow, 
we numerically integrate the equations of motion (\ref{eq:motion_j}) for the four vortices, $j=1,\dots,4$, varying  
$r/R$ and $R/D$. In particular, we choose $r/R=n/10$ and $R/D=m/10$, with $m, n =1, \dots , 9$. The time-advancement scheme employed 
in the numerical simulations is a second-order Adams-Bashforth method with a time step $\Delta t=T/1000$ where $T=2\pi^2\delta^2/\kappa$
is the rotation period of a pair of vortices of the same polarity placed at distance $\delta$. In our numerical simulations $\delta$ is set 
to $10^{-3}D$. 

For classical unbounded fluids, since the study performed by 
Love over a century ago \citep{Love1894}, it is well known that vortices
undergo leapfrogging motion only if $r/R$ is larger than a critical 
value $\alpha_c = 3 - 2\sqrt{2} \approx 0.172$. If $r/R < \alpha_c$,
leapfrogging does not occur: the smaller, faster pair moves
``too fast'' for the larger ring to influence its dynamics in a significant
way, and the vortices separate.
More recently, Acheson \citep{Acheson2000} extended numerically the study performed by Love and established
that leapfrogging motion is unstable when $\alpha_c < r/R < \alpha_c'$, with $\alpha_c'=0.382$. 

\begin{figure}
\begin{center}
\includegraphics[width=0.9\textwidth]{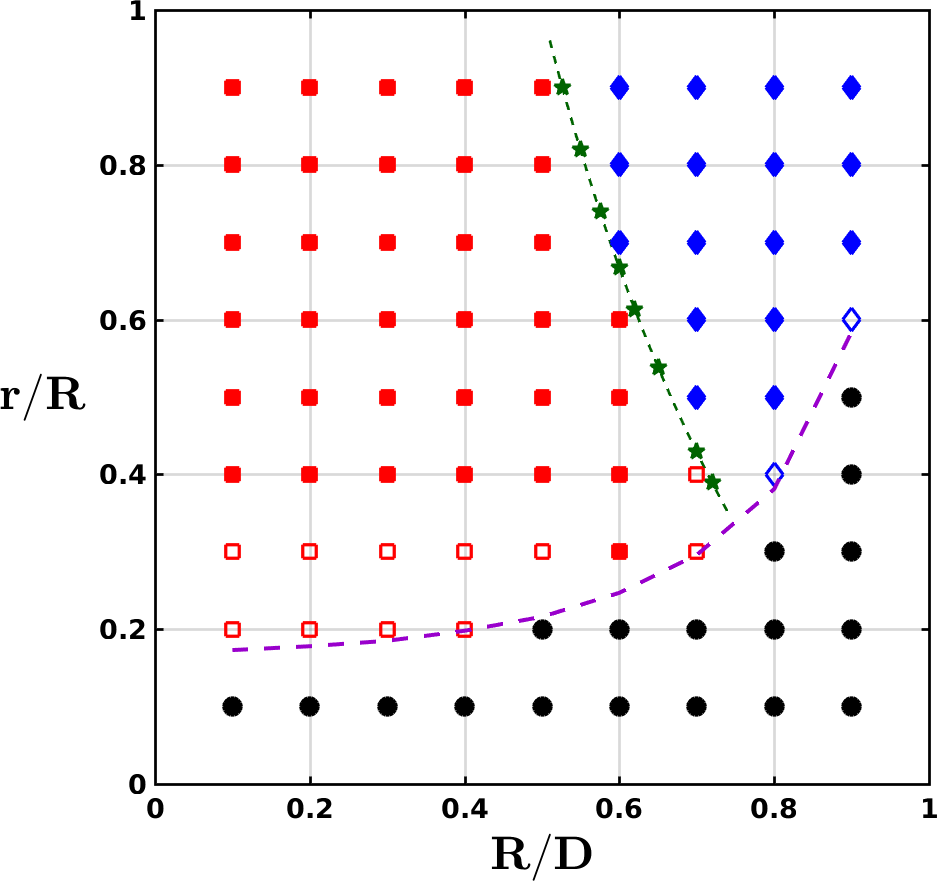}
\caption{Phase diagram of the classical motion of two vortex-anti vortex pairs in a two-dimensional plane channel. All symbols 
refer to performed numerical simulations. Black circles indicate 
no leapfrogging motion; red filled (open) squares stand for stable (unstable) forward, standard leapfrogging; blue filled (open) 
diamonds correspond to stable (unstable) backward leapfrogging; green stars stand for periodic orbits. The dashed green line indicates
the analytical solution for periodic orbits (see section \ref{subsebsec:deriv_part1} and Appendix). The dashed violet curve is the numerically computed $\alpha_c$ dependence on
$(R/D)$.}
\label{fig:phase.diag}
\end{center}
\end{figure}

In our two-dimensional channel, the confinement of the flow leads 
to a richer dynamics than in an unbounded domain. 
In addition to distinction between leapfrogging and non-leapfrogging,
which is already known, we also observe 
{\it backward leapfrogging} and {\it periodic orbits}. 
The phase diagram of the system resulting from the numerical simulations 
is illustrated in Fig.~\ref{fig:phase.diag}.

For values of $R/D \le 1/2$,
the dynamics is very similar to what is observed in an unbounded fluid, 
the role of the boundaries being only marginal. 
For a given value of $R/D \le 1/2$, in fact, as we increase
$r/R$, we first observe non leapfrogging motion 
(in black in Fig.~\ref{fig:phase.diag}), defined as the dynamics 
characterised by $\dot{y}_j(t)=0$ for all $j$ at late times; 
then we notice unstable leapfrogging motion (open red squares), 
and finally stable leapfrogging (filled red squares). 
These dynamical regimes therefore
coincide with the scenario outlined by 
Acheson \citep{Acheson2000}, the only significant and important 
difference being the dependence of $\alpha_c$ on $R/D$: for small 
values of $R/D$, $\alpha_c$ is very close 
to the constant value $0.172$ for vortex leapfrogging in 
unbounded fluids ({\it e.g.} for $R/D=0.1$, $\alpha_c=0.173$), 
increasing for
increasing values of $R/D$ ({\it e.g.} $\alpha_c=0.216$ for $R/D=0.5$). 
This dependence of $\alpha_c$ on $R/D$ stems from the 
interaction
of the outer vortices 1 and 2 in Fig.~\ref{fig:ic_classical}) 
with their corresponding images with respect to the closest 
channel wall; essentially, the interaction with image vortices 
is stronger compared to the interaction of the inner pair 
with the corresponding images. 
These images, of opposite sign, slow down the outer vortex pair, 
allowing the inner pair to escape towards infinity for values of $r/R$
which would produce leapfrogging motion in an unbounded fluid; 
in order to recover leapfrogging, $r/R$ would have to increase. 
As $R/D$ increases, this effect is amplified as the outer pair 
is closer to the channel walls.

\begin{figure}
\begin{center}
\begin{minipage}{0.48\textwidth}
      \centering
       \includegraphics[width=0.95\textwidth]{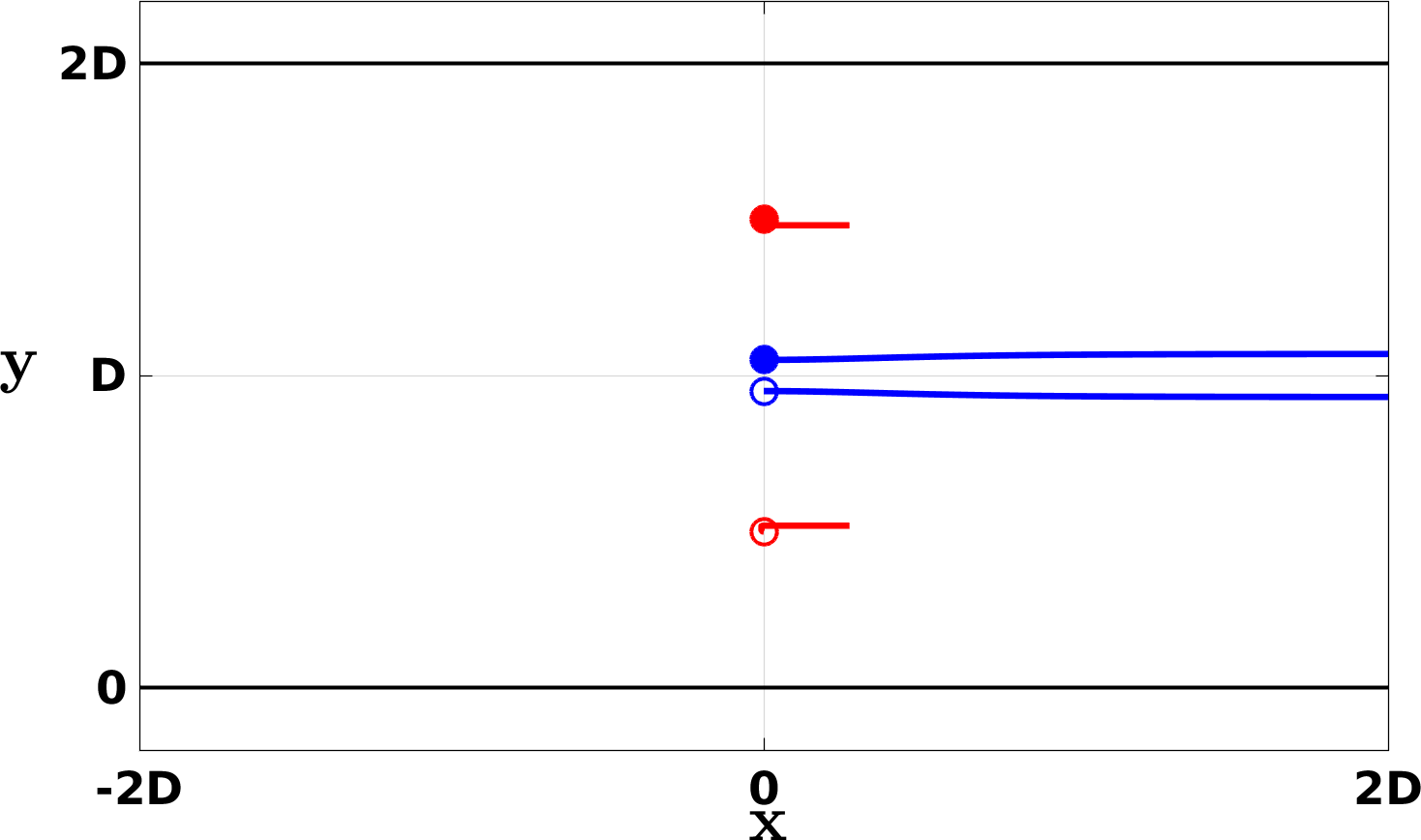}
\end{minipage}
\hspace{0.01\textwidth}
\begin{minipage}{0.48\textwidth}
      \centering
       \includegraphics[width=0.95\textwidth]{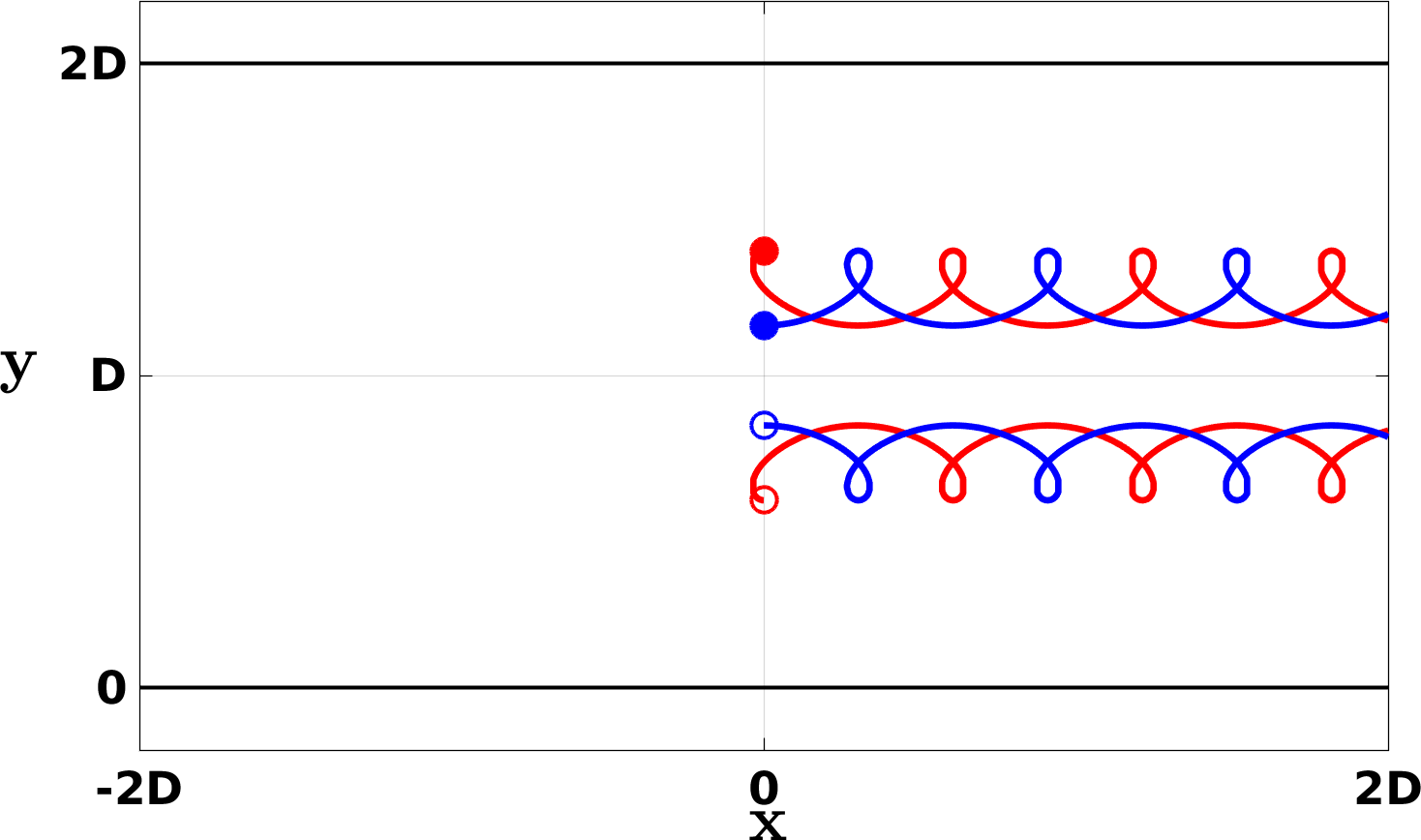}
\end{minipage}\\[5mm]
\begin{minipage}{0.48\textwidth}
      \centering
       \includegraphics[width=0.95\textwidth]{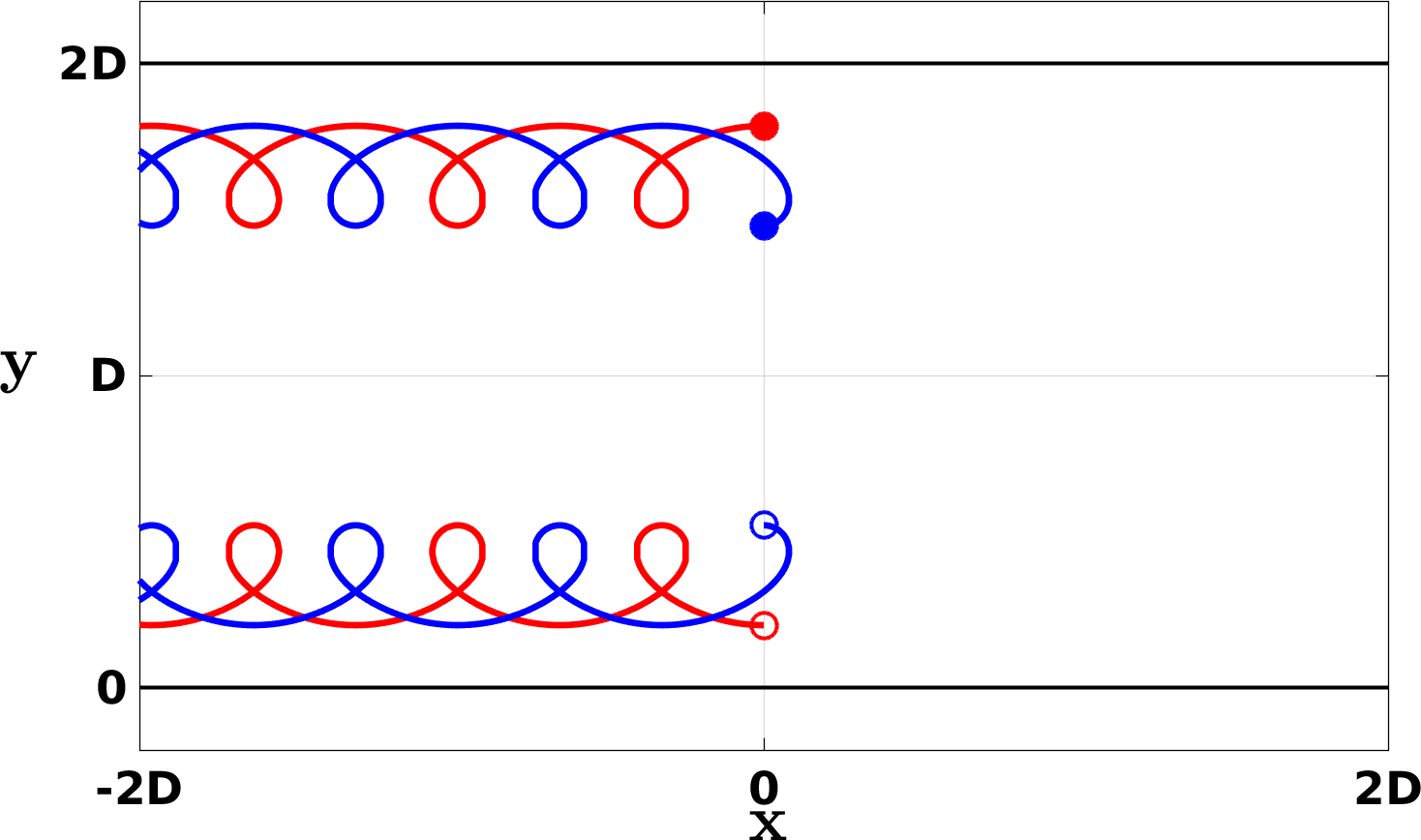}
\end{minipage}
\hspace{0.01\textwidth}
\begin{minipage}{0.48\textwidth}
      \centering
       \includegraphics[width=0.95\textwidth]{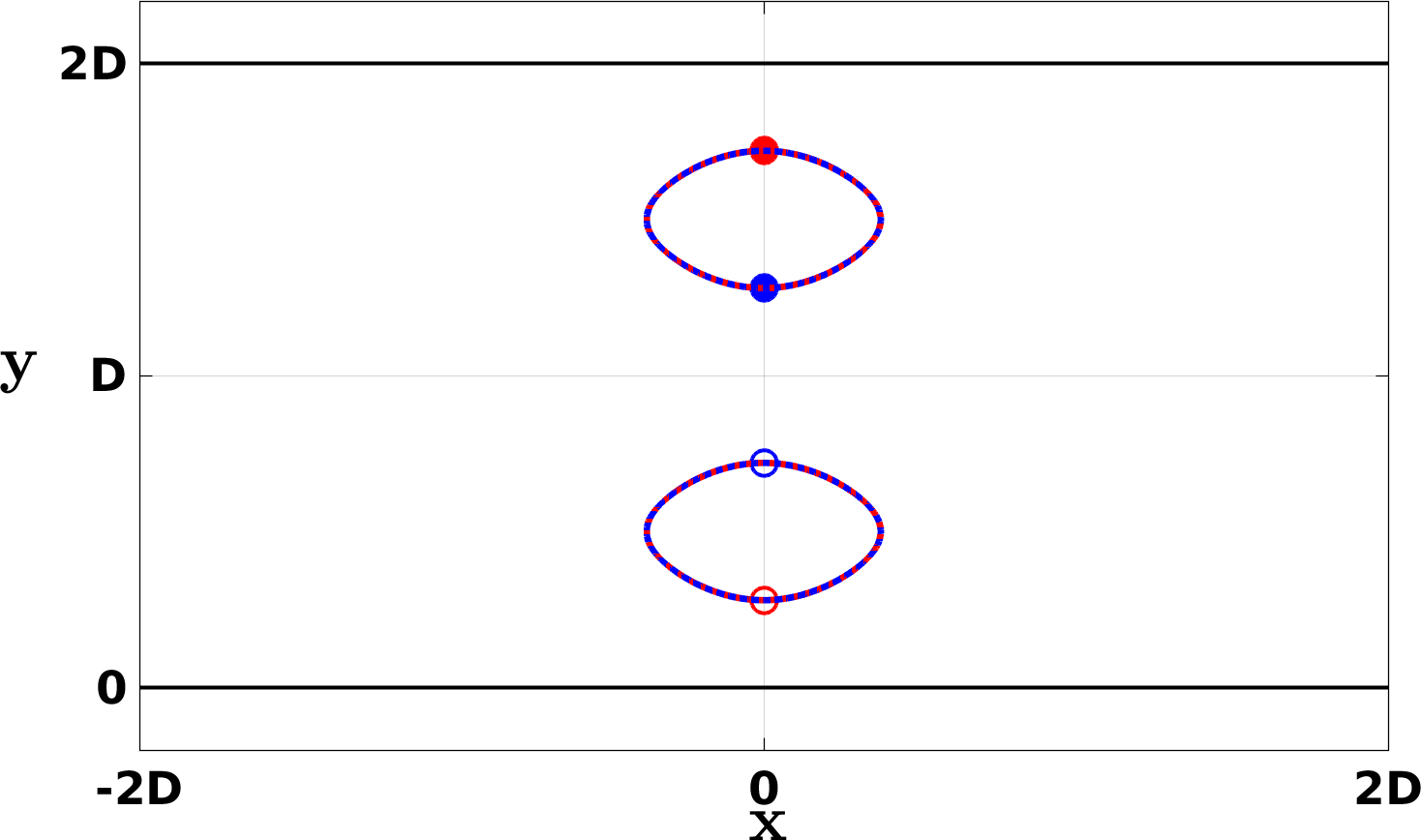}\\[1mm]
\end{minipage}
\caption{
Examples of dynamical regimes and trajectories for classical 4-vortex motion in a two-dimensional channel. 
Filled (open) symbols indicate positive (negative) vortices. 
Top Left: 
$R/D=5/10$, $r/R=1/10$, no-leapfrogging 
(vortices moving to the right). 
Top Right: 
$R/D=4/10$, $r/R=4/10$, forward (standard) leapfrogging
(vortices moving to the right). 
Bottom Left: 
$R/D=8/10$, $r/R=6/10$, backward-leapfrogging 
(vortices moving to the left); 
Bottom Right: $R/D=72/100$, $r/R=39/100$, periodic orbits.}
\label{fig:all.dyamics}
\end{center}
\end{figure}

This increasing monotonous behaviour of $\alpha_c$ with respect to
$R/D$ extends also for $R/D>1/2$, where the role played by boundaries becomes 
significant, triggering a much richer dynamics.
As $R/D$ is larger than $1/2$, for large values of $r/R$, 
we observe {\it backward-leapfrogging}, indicated by blue diamonds in 
Fig.~\ref{fig:phase.diag}. This dynamics, again, originates 
from the interaction of vortices with their images with respect to
the closest channel wall. In particular, each vortex, paired to
its image of opposite sign, forms a {\it virtual} vortex-anti 
vortex pair on its own. 
As a consequence, we observe two distinct leapfrogging motions, 
each involving two virtual vortex-anti vortex pairs. 
Due to the vortex polarity, the leapfrogging motion induces a net
translation in the opposite direction with respect to standard (forward)
leapfrogging. 
In the $(R/D,r/R)$ plane, the forward-leapfrogging to backward-leapfrogging 
transition occurs via an intermediate regime in which vortices
follow periodic orbits, indicated by green stars 
in Fig.~\ref{fig:phase.diag}. As shown in detail in the next section 
and in the analytical derivation presented in the Appendix,
periodic orbits are observed when $R + r =D$, corresponding to the green dashed line in Fig.~\ref{fig:phase.diag}. 
For large values of $R/D$ ($R/D\gtrsim 3/4$), 
the system crosses directly the no-leapfrogging to backward-leapfrogging 
boundary without
passing through a forward-leapfrogging regime. 
Examples of all the different regimes observed in our 
system of classical point vortices are shown in Fig.~\ref{fig:all.dyamics}.
Note that in the three-dimensional coaxial vortex rings analog, 
vortices of the same colour correspond to cross-sections 
of the same vortex ring.

\subsubsection{Derivation of periodic orbits}
\label{subsebsec:deriv_part1}

In this section  
we derive theoretically the existence of periodic orbits in the 
leapfrogging motion of four vortices in a channel using
the classical point-vortex model. 
We show that under suitable conditions, 
namely when $R+r=D$, each pair of same signed vortices moves 
around a fixed point. Some analytic details are discussed in the Appendix.
%

With reference to Fig.~\ref{fig:ic_classical}, we consider the pair of vortices $P_1 =(x_0(t),D-R(t))$, with negative circulation
$-\kappa$, and  $P_2= (x_0(t),D+R(t))$, with positive circulation
$\kappa$, and the pair of vortices $P_3 =(x_1(t),D-r(t))$, with negative circulation
$-\kappa$, and  $P_4= (x_1(t),D+r(t))$, with positive circulation
$\kappa$, where $t$ is time. 
In the complex domain, omitting the time dependence to ease notation,  
these vortices are located in $z_1 =x_0 +i (D-R)$ for $P_1$, 
$z_2 =x_0 +i (D+R)$ for $P_2$, $z_3 =x_1 +i (D-r)$ for $P_3$ and 
$z_4 =x_1 +i (D+r)$ for $P_4$, and they generate the following complex 
velocity in the point $z$, as given by Eq. \eqref{eq:cplx_vel_N},

\begin{equation}
\displaystyle
w(z)=w(z,z_1)+w(z,z_2)+w(z,z_3)+w(z,z_4)  \;\; .
\label{eq:cplx_vel_N4}
\end{equation}

\noindent
We now consider the midpoint $M$ between the vortex points $P_1$ and $P_3$, 
namely  
$\displaystyle  z_M(t)~=~\frac{x_0(t)+x_1(t)}{2}+i\left(D-\frac{r(t)+R(t)}{2}\right)$ 
and the complex velocity generated by vortices in $z_M$ which we 
indicate with $w(z_M)$

\begin{equation}
\displaystyle
w(z_M)= \frac{\frac{i \kappa}{2 D} \left(-1+e^{\frac{2 i \pi  (r+R)}{D}}\right) e^{\frac{\pi 
   (\text{x}_0+\text{x}_1)}{2 D}}}{ \left(e^{\frac{\pi  (4 i r+4 i
   R+\text{x}_0+\text{x}_1)}{2 D}}-e^{\frac{\pi  (2 \text{x}_0+i (r+3 R))}{2
   D}}-e^{\frac{\pi  (2 \text{x}_1+i (3 r+R))}{2 D}}+e^{\frac{\pi 
   (\text{x}_0+\text{x}_1)}{2 D}}\right)}  \;\; .
\label{eq:cplx_vel_zM}
\end{equation}

\noindent
If we look for the conditions such that the velocity $w(z_M)$ of the midpoint $M$ is zero, we have

\begin{equation}
\displaystyle
w(z_M)= 0 \quad  \Longleftrightarrow \quad e^{\frac{2 i \pi  (r+R)}{D}}-1=0 \quad  \Longleftrightarrow \quad 
\frac{2\pi  (r+R)}{D} =  2k \pi \; , \quad k\in \mathbb{Z} \;\;. 
\label{eq:cplx_vel_zM_zero}
\end{equation}

\noindent
Note that the same result Eq. \eqref{eq:cplx_vel_zM_zero} is found for the 
midpoint $N$ between the two vortex points $P_2$ and $P_4$.

Since $r$, $R$, and $D$ are positive real parameters, the only 
admissible values of 
$k$ in \eqref{eq:cplx_vel_zM_zero}   are $k \in \mathbb{Z}^+$. Moreover, 
we know that $r < R < D$, leading to $r + R < 2D$, which implies that 
the only admissible value for $k$ is $k=1$, \textit{i.e.} 

\begin{equation}
\displaystyle
  r(t)+R(t)=D \;\;.
\label{eq:cplx_vel_zM_zero_con}
\end{equation}

\noindent
This is the most interesting result: it states that when the four vortices satisfy the condition \eqref{eq:cplx_vel_zM_zero_con} then 
the midpoints $M$ and $N$ are at rest: the two pairs of vortices $(P_1,\, P_3 )$ and $(P_2,\, P_4 )$ move hence symmetrically with respect to 
their correspondent midpoints, \textit{i.e.} $\dot{x}_0(t)=-\dot{x}_1(t)$ and $\dot{R}(t)=-\dot{r}(t)$. The last equality is fundamental as
it expresses that if condition \eqref{eq:cplx_vel_zM_zero_con} is satisfied at a given $t=t_0$, it will be satisfied for every $t>t_0$.
Thus, if the initial condition is prepared such that $x_0(0)=x_1(0)=0$ and $r(0)+R(0)=D$, vortices will always move symmetrically with
respect to their midpoints $\displaystyle  z_M=i \frac{D}{2}$  
and $\displaystyle  z_{N}= i \frac{3D}{2}$.
%

The last step to demonstrate the existence of periodic orbits 
is to prove that the trajectories of the 
vortex points are closed curves rotating around the two midpoints 
$M$ and $N$ as, in principle more general trajectories   
with the restriction $\dot{R}(t)=-\dot{r}(t)$ (for instance, $\dot{R}(t)=\dot{r}(t)=0$) 
could be  possible, not leading to periodic orbits. 
We tackle this issue in the Appendix, to ease the readability of the
manuscript.

\subsection{Quantum fluids}
\label{sec:results.quantum}

The next step is to numerically probe the dynamical regimes of 
two quantum vortex-antivortex pairs interacting in a two-dimensional 
channel. We shall compare the results
with the corresponding classical results outlined 
in the previous Section~(\ref{sec:results.classical}).

We consider a two-dimensional BEC 
in a channel geometry, imprinting quantum vortices in the 
positions initially occupied by classical vortices.
Note that, in addition to the parameters
$R$, $r$ and $D$ already present in the classical point vortex
formulation, in the Gross-Pitaevskii formulation of the problem
we have an extra length scale - the healing length $\xi$ - which plays
a fundamental role in the dynamics.
To assess the relevance of this extra length scale, 
we present numerical simulations of leapfrogging quantum vortices
employing two distinct values of the channel half-width $D$:
$D_1=40\xi$ and $D_2=20\xi$. In order to model
the channel confinement, we use the following potential $V$:

\begin{equation}
V = V(y) = 
  \begin{cases} 
   0      & \text{if } 0 < y <  2D \\[3mm]
   10\mu  & \text{if } \; y \le 0 \;\; \text{or } \; y \ge 2D , 
  \end{cases}
  \label{eq:channel_potential}
\end{equation}

\noindent
corresponding to a channel of half-width $D$, 
where the density $|\Phi|^2$ is constant everywhere with the exception of
thin layer whose width is of the order of the healing length
at the channel boundaries $y=0$ and $y=2D$. 

The trajectories of the quantum vortices 
are calculated as a function of time by numerically solving the
equation of motion of the order parameter $\Phi$,
the dimensionless Gross-Pitaevskii equation

\begin{equation} 
\displaystyle
i\dot{\Phi} = -\frac{1}{2}\Delta \Phi + \frac{V}{\mu}\Phi + |\Phi|^2 \Phi  - \Phi\;\; .
\label{eq:GP.non.dim}
\end{equation}

\noindent
Equation~(\ref{eq:GP.non.dim}) is obtained from Eq.~(\ref{eq:GPmu}) 
after introducing characteristic units of length,
time and energy: $\xi=\hbar/\sqrt{m\mu}$ (the healing length),  
$\tau=\xi/c\;$ (where $c=\sqrt{\mu/m}$ is the speed of sound),
and $\mu$ (the chemical potential) respectively, and 
normalising the order parameter with respect to the unperturbed 
homogeneous solution
$\Phi_0=\sqrt{\mu/g}$ of Eq.~(\ref{eq:GPmu}).
In these units the healing length and the bulk density in the channel
are unity.

The numerical integration of Eq. (\ref{eq:GP.non.dim}) is performed employing a fourth-order Runge-Kutta time advancement scheme
and second-order finite differences to approximate spatial derivative operators. 
Time step $\Delta t/\tau$ is set to $1.5\times 10^{-2}$ and
spatial discretization $\Delta x/\xi = \Delta y/\xi$ is chosen to be equal to $0.25$. 
In the set of simulations where $D=D_1=40\xi$, the numbers of grid-points in the $x$ and $y$ 
directions are $N_x=6400$ and $N_y=400$ respectively, leading to the
computational box $-800\xi \le x \le 800\xi$ and $-10\xi \le y \le 90\xi$. On the other hand, when $D=D_2=20\xi$, 
$N_x=3200$ and $N_y=240$ respectively, leading to the
computational box $-400\xi \le x \le 400\xi$ and $-10\xi \le y \le 50\xi$.

\begin{figure}
\begin{center}
\begin{minipage}{0.48\textwidth}
      \centering
       \includegraphics[width=0.95\textwidth]{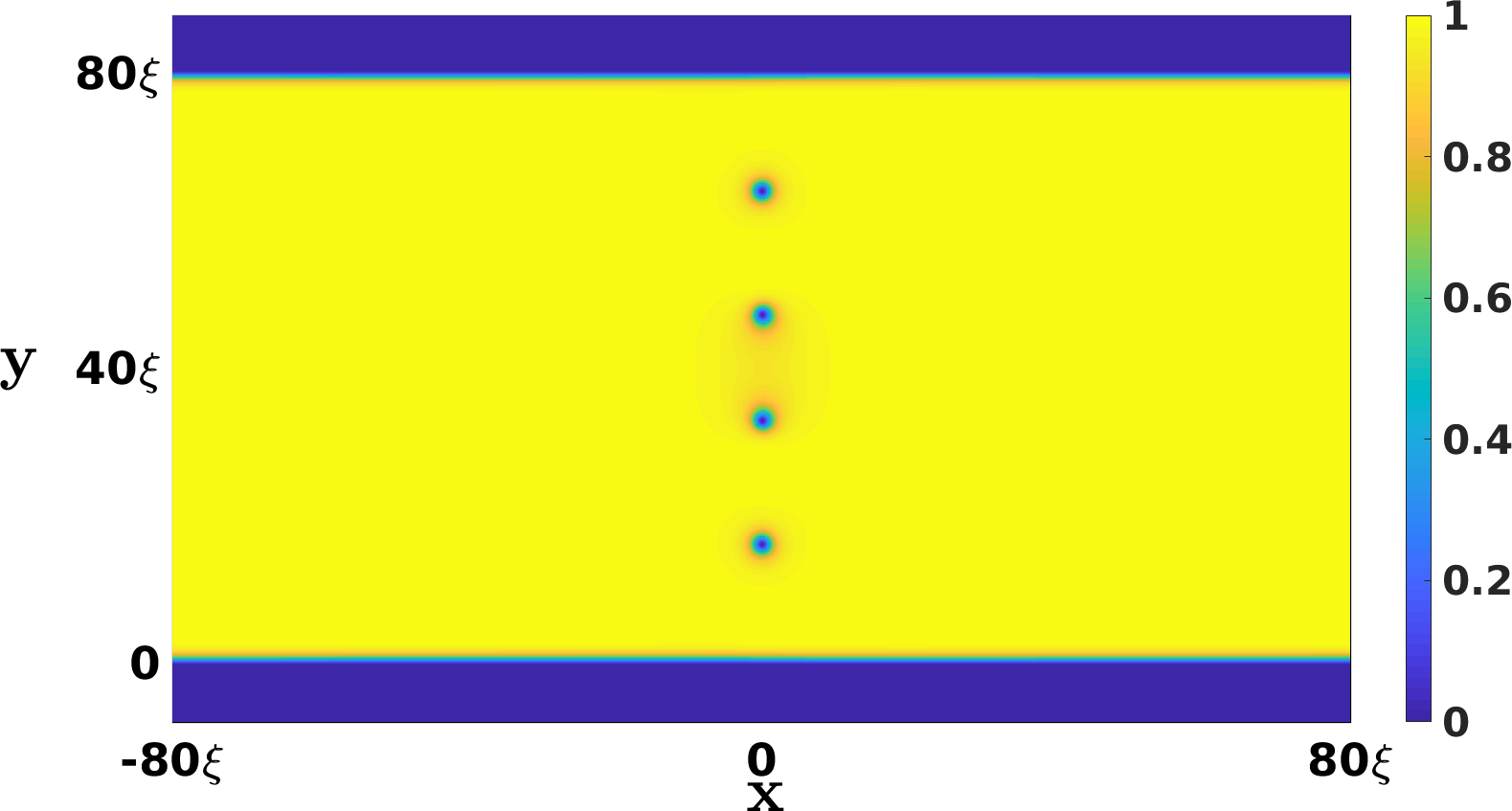}
\end{minipage}
\hspace{0.01\textwidth}
\begin{minipage}{0.48\textwidth}
      \centering
       \includegraphics[width=0.95\textwidth]{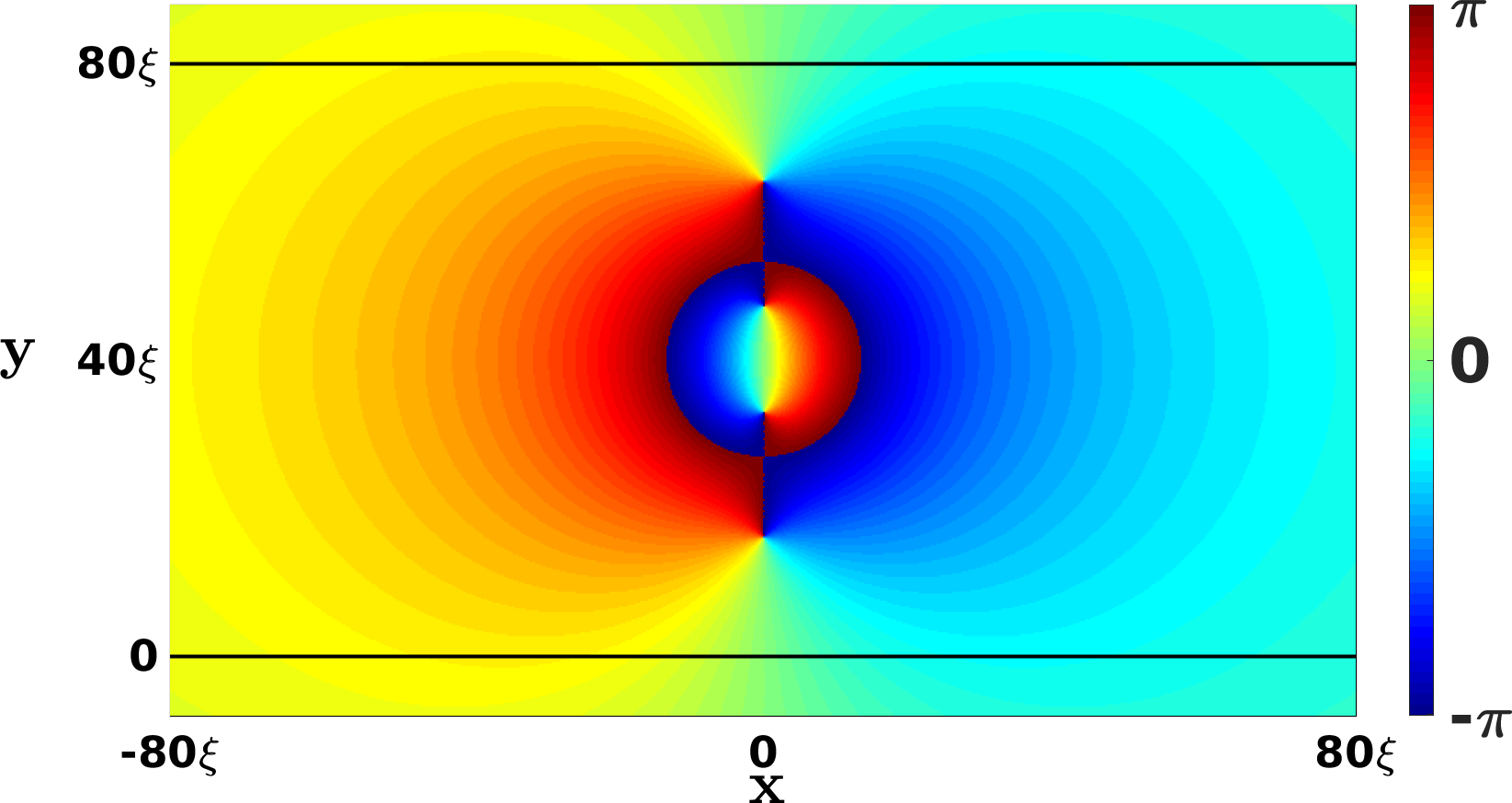}
\end{minipage}
\caption{Initial condition for numerical simulation of leapfrogging 
of quantum vortices in a 
two dimensional channel for $R/D=0.6$ and $r/R=0.3$ and $D=D_1=40\xi$.
(Left) the density of the BEC $|\Phi(x,y)|^2$ (presented as a ratio of the bulk density $|\Phi_0|^2$) is displayed:
it is unity (yellow) in the bulk of the channel and vanishes (blue)
in the vortex cores and at the channel's boundaries;
(right) the phase $\theta(x,y)$ of the BEC is illustrated in 
the range $[-\pi,\pi )$.}
\label{fig:ic_quantum}
\end{center}
\end{figure}

The initial imprinting of vortices is made by enforcing a uniform $2\pi$ phase wrapping
around the positions employed as initial condition for the classical point vortex simulations 
and letting the system relax in imaginary time before starting the 
integration of Eq. (\ref{eq:GP.non.dim}) for $t\in \mathbb{R}$. In Fig.~\ref{fig:ic_quantum} we report the density $|\Phi|^2(x,y)$ (left)
and the phase $\theta(x,y)$ (right) of the initial condition employed for $R/D=0.6$ and $r/R=0.3$ and $D=D_1=40\xi$. 
It can be easily observed that the density
$|\Phi|^2$ rapidly drops to zero  at the vortex positions and outside
the channel. Correspondingly, the four $2\pi$ phase wrappings can be distinguished in Fig.~\ref{fig:ic_quantum} (right).

To verify the existence in a BEC of all distinct regimes 
observed in the classical point vortex model 
(Section \ref{sec:results.classical}), 
we perform numerical simulations of quantum vortex leapfrogging 
along the vertical line $R/D=0.6$ of
the phase-diagram reported in Fig.~\ref{fig:phase.diag}; 
we have chosen this value of $R/D$ because along this line, as
$r/R$ varies from $1/10$ to $9/10$, all regimes which we have identified
using the classical point vortex model are present.

\begin{figure}
\begin{center}
\includegraphics[width=0.9\textwidth]{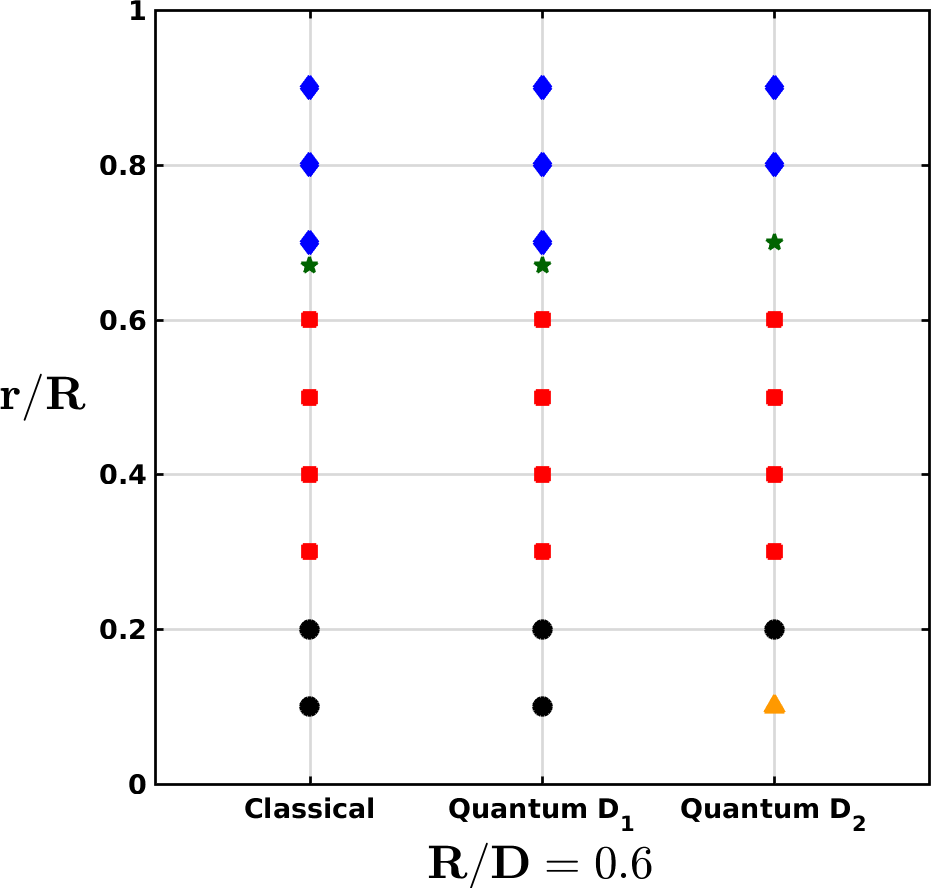}
\caption{Cuts in the dynamical regimes phase diagram corresponding to $R/D=0.6$ 
for the classical point vortex model (left), the Gross-Pitaevskii model with $D=D_1=40\xi$ (middle) and 
$D=D_2=20\xi$ (right). Symbols as in Fig~\ref{fig:phase.diag} except for the newly introduced 
up-pointing orange triangle corresponding to the
annihilation of the inner vortex-anti vortex pair.}
\label{fig:phase.diag_quantum}
\end{center}
\end{figure}
 
The results are schematically outlined in Fig.~\ref{fig:phase.diag_quantum}, 
where classical vortex dynamics (left)
is compared to quantum vortex dynamics 
at $D=D_1=40\xi$ (middle) and $D=D_2=20\xi$ (right). 
When $D=D_1$, the boundaries of the phase diagram at $R/D=0.6$
are at the same values of $r/R$ in the classical and in the quantum case.
When $D=D_2$ we observe two differences: first, periodic motion
now occurs at $(R/D,r/R)=(0.6\, ,\,0.7)$ instead of
$(R/D,r/R)=(0.6\, ,\,0.67)$; second, at 
$(R/D,r/R)=(0.6\, ,\,0.1)$ the internal vortex-anti vortex pair 
annihilates as their initial distance is only $2.4\xi$.
These differences at the smaller value of channel size are expected,
as the healing length scale starts playing a role: only if $D/\xi$ is 
sufficiently large we can expect classical and quantum dynamics
to be the same.

\begin{figure}
\begin{center}
\begin{minipage}{0.48\textwidth}
      \centering
       \includegraphics[width=0.95\textwidth]{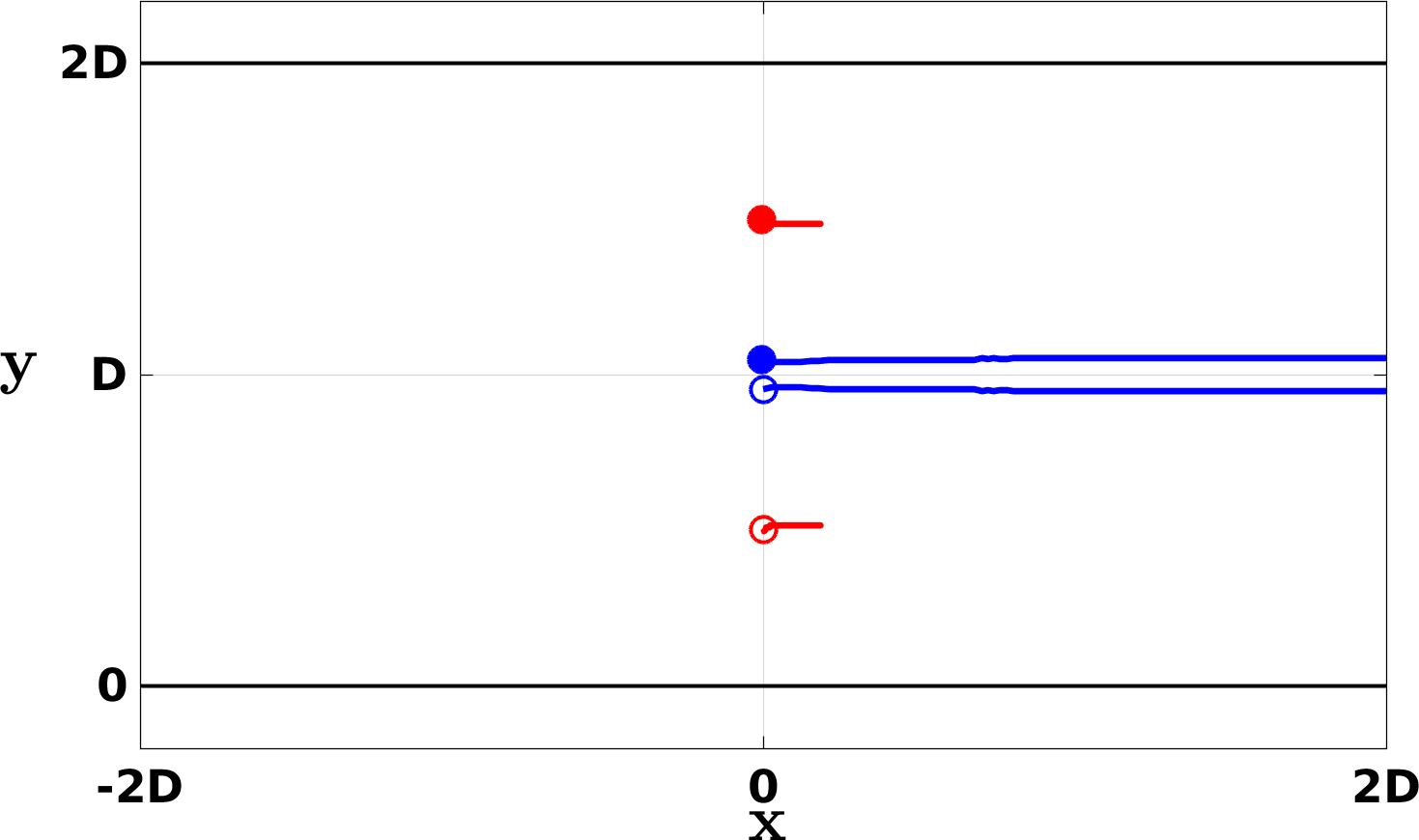}
\end{minipage}
\hspace{0.01\textwidth}
\begin{minipage}{0.48\textwidth}
      \centering
       \includegraphics[width=0.95\textwidth]{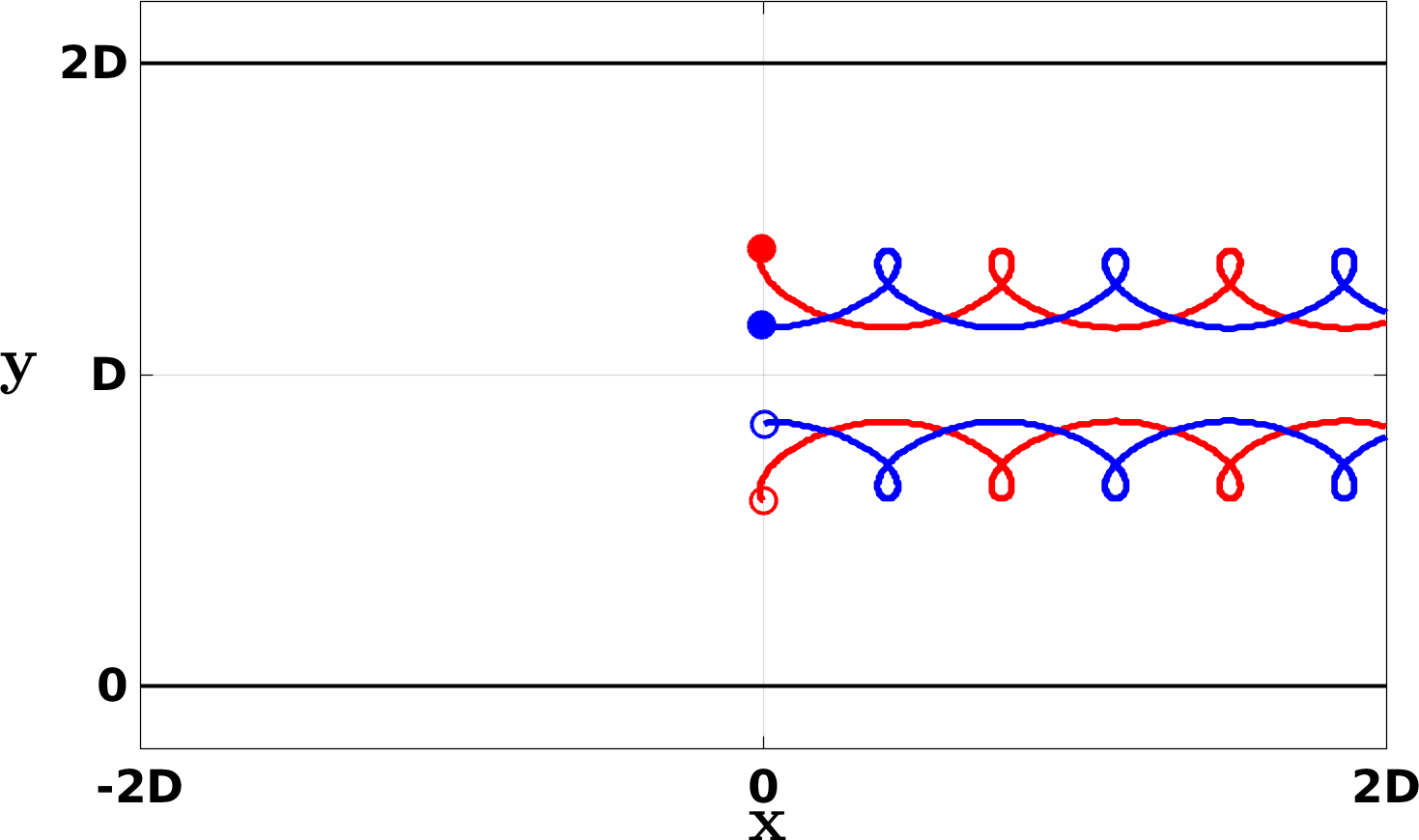}
\end{minipage}\\[5mm]
\begin{minipage}{0.48\textwidth}
      \centering
       \includegraphics[width=0.95\textwidth]{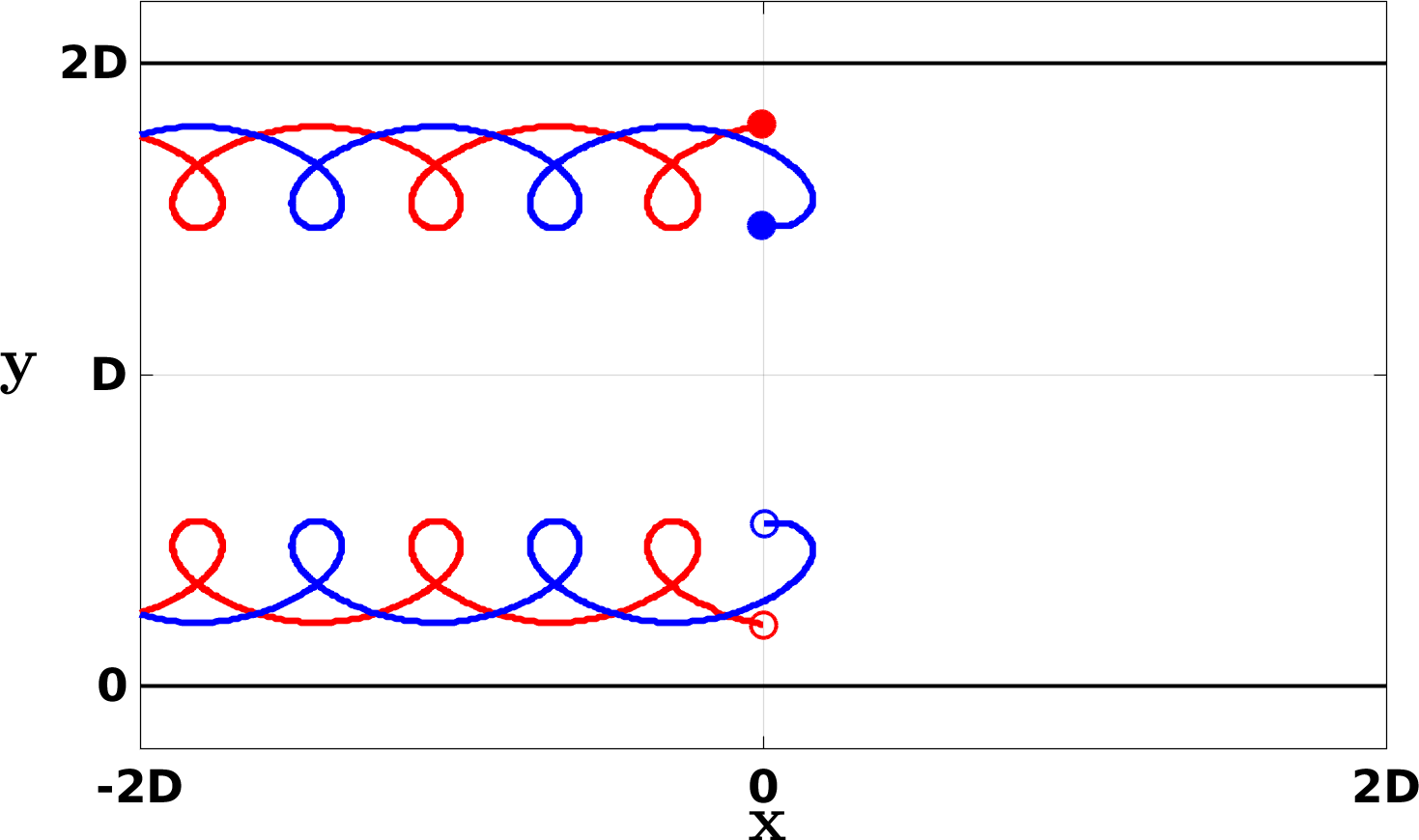}
\end{minipage}
\hspace{0.01\textwidth}
\begin{minipage}{0.48\textwidth}
      \centering
       \includegraphics[width=0.95\textwidth]{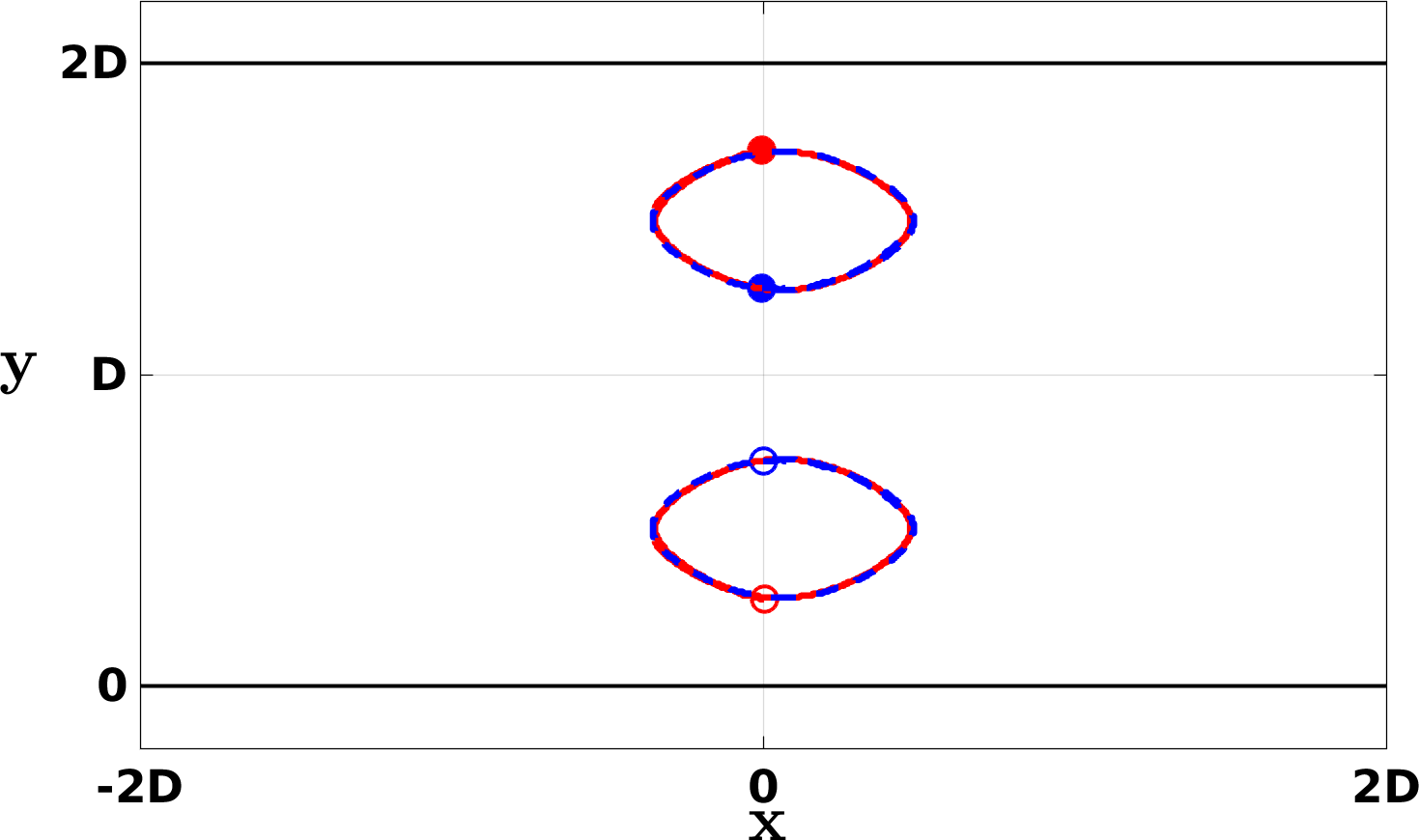}\\[1mm]
\end{minipage}
\caption{Dynamical regimes observed in a quantum 4-vortex configuration 
in a two-dimensional channel. 
Filled (open) symbols
indicate positive (negative) vortices. Top Left: $R/D=5/10$, $r/R=1/10$, no leapfrogging motion is observed. Top Right: 
$R/D=4/10$, $r/R=4/10$, forward (standard) leapfrogging. 
Bottom Left: $R/D=8/10$, $r/R=6/10$, backward-leapfrogging; 
Bottom Right:
$R/D=72/100$, $r/R=39/100$, periodic orbits.}
\label{fig:all.dyamics.quantum}
\end{center}
\end{figure}

The exact matching of the observed dynamical regimes when comparing 
classical and quantum leapfrogging in a two-dimensional channel 
if $D\ge 40\xi$ is confirmed in Fig.~\ref{fig:all.dyamics.quantum},
which shows the trajectories of quantum vortices for 
$(R/D \, , r/R)$ pairs selected as for 
the classical trajectories illustrated in Fig.~\ref{fig:all.dyamics}.

It is worth noting some minor differences between the 
quantum vortex trajectories and their classical counterparts
reported in Fig.~\ref{fig:all.dyamics}. Since the initial
condition is not stationary with respect to any frame of reference, 
when we start integrating in time Eq.~(\ref{eq:GP.non.dim}) 
for $t\in \mathbb{R}$ there is a sudden emission of sound waves,
and as a result the entire vortex configuration is translated 
towards the positive $x$ direction.  The effect (which has been
reported in the literature \citep{frisch-etal-1992}),
is visible in the top right, bottom left and bottom right panels of 
Fig.~\ref{fig:all.dyamics.quantum} when compared Fig.~\ref{fig:all.dyamics}.
In particular, this horizontal shift affects
the periodic orbits reported in Fig.~\ref{fig:all.dyamics.quantum} 
(bottom right) whose center is slightly shifted towards positive $x$ values. 
In addition, the number of 
periods observed in the $x$ range $[-2D, \, 2D]$ is different from
the classical counterpart, possibly due to the 
compressible nature of a quantum Bose gas, in which
incompressible kinetic energy may be transformed into compressible 
kinetic energy (sound) when vortices change their velocities
(accelerate), as shown by \citet{Parker2004}, exactly as the accelerated 
motion of charged particles emit electro-magnetic radiation. 
The role played by this effective dissipation of kinetic energy into
sound will be assessed in a future study.

\section{Conclusions}
\label{sec:concl}

In conclusion, we have demonstrated that, in the confined space of a two
dimensional channel, the classical problem of vortex leapfrogging acquires
new aspects. Using the point vortex model we have found that,
besides the known regimes of standard leapfrogging and
absence of leapfrogging, there are two new regimes: backward leapfrogging 
and periodic motion.  
Using the Gross-Pitaevskii equation to model an atomic Bose-Einstein condensate 
(a compressible quantum fluid) confined within a channel, we have verified that all four 
regime also exist for quantum vortices. 
In large channels, the boundaries
between these regimes are the same for classical and quantum vortices.
Some differences appear if the channel size is reduced, and
the finite-size nature of the quantum
vortex core starts playing a role, or if the vortices are very close
and sound radiation becomes important. The determination of a 
richer dynamics for the leapfrogging of vortices occuring in confined geometries
will be particularly important for the interpretation and planning
of ongoing and future experiments with atomic Bose-Einstein Condensates,
where the dynamical regimes reported in the present work can be potentially
observed.

Future work will address the problem in three dimensions,
paying attention to the excitation of Kelvin waves along the
vortex rings and the departure from axisymmetry.

\section{Acknowledgements}

LG, NGP and CFB acknowledge the support of the Engineering and Physical Sciences 
Research Council (grant No.  EP/R005192/1).
M.S. acknowledges the support of MIUR-Italy through the project PRIN ``Multiscale phenomena in Continuum Mechanics: singular limits, off-equilibrium and transitions" (grant No. PRIN2017 2017YBKNCE).

Declaration of Interests. The authors report no conflict of interest.




\appendix

\section{Derivation of periodic orbits}
In order to show the existence of periodic orbits, we have to prove that if condition \eqref{eq:cplx_vel_zM_zero_con} is satisfied, 
the trajectories of the vortex points are closed curves with vortices rotating  around the two midpoints $M$ and $N$ defined in 
section \ref{subsebsec:deriv_part1}.

For the sake of simplicity, and with reference to section \ref{subsebsec:deriv_part1},
we prove the closedness of the trajectory only for the vortex point $P_1$, as the proof 
for the other vortex points is an iterative procedure.
We consider equation \eqref{eq:cplx_vel_j} for the vortex point $z_1$ with the complex velocity given by the 
expression \eqref{eq:cplx_vel_N4} evaluated on the vortex point $z_1$. 
Since the middle point $z_M$ is at rest for $r+R=D$, 
we rewrite the dynamic equation of $z_1$, namely $\dot{z_1}= w^*(z_1)$, 
in the polar coordinate system $(\rho,\, \theta)$ centered on $z_M$. The   middle point $z_M$,  under  the condition $r+R=D$, becomes $\displaystyle  z_M~=~\frac{x_0+x_1}{2}+i\frac{D}{2}$, which requires the condition $\displaystyle \rho< \frac{D}{2}$ to ensure that vortices $P_1$ and $P_3$ are in $\{ z\in\mathbb{C} : 0 < \Im\text{m} \; z < D \}$.  

  Thus, in the new reference system the vortex points correspond to
\begin{equation}\label{coord_points}
 \begin{array}{ll}
 z_1= z_M-\rho \cos(\theta)-i \rho \sin(\theta), \\
 z_2= z_M+iD-\rho \cos(\theta)+i \rho \sin(\theta), \\
 z_3= z_M+\rho \cos(\theta)+i \rho \sin(\theta), \\
 z_4= z_M+iD+\rho \cos(\theta)-i \rho \sin(\theta), \\
 \end{array}
\end{equation} 
where $z_M$ is now the origin of the new frame of reference, which can be set  $z_M=0 + 0 i$. 
Note that the condition  \eqref{eq:cplx_vel_zM_zero_con} is automatically satisfied by construction; 
indeed, $\displaystyle \frac{z_2-z_1}{2}~=~i~\left(\frac{D}{2}+\rho\sin(\theta)\right)$ and 
$\displaystyle \frac{z_4-z_3}{2}~=~i~\left(\frac{D}{2}-\rho\sin(\theta)\right)$, 
implying  $\displaystyle R~\equiv~\frac{D}{2}+\rho\sin(\theta)$ and $\displaystyle r~\equiv~\frac{D}{2}-\rho\sin(\theta)$ 
and, hence, condition \eqref{eq:cplx_vel_zM_zero_con}. 
We now substitute the coordinates \eqref{coord_points} into the equation
\begin{equation}\label{equa:vort_point_z1}
\dot{z}_1=w^*(z_1,z_{k_{\{k=1,\dots,4\}}}),
\end{equation}
according to \eqref{eq:cplx_vel_j}, and change the vectorial basis from $(\hat{\mathbf{x}},\, \hat{\mathbf{y}})$ to 
$(\hat{\mathbf{u}}_\rho,\, \hat{\mathbf{u}}_\theta)$  by means of the following rotation:
\[
\hat{\mathbf{u}}_\rho =\cos(\theta) \hat{\mathbf{x}} + \sin(\theta) \hat{\mathbf{y}}, 
\qquad \hat{\mathbf{u}}_\theta= -\sin(\theta) \hat{\mathbf{x}} + \cos(\theta) \hat{\mathbf{y}}. 
\]
By writing $\dot{z_1}=-\dot{\rho} \hat{\mathbf{u}}_\rho-\rho \dot{\theta} \hat{\mathbf{u}}_\theta$, 
we then find the following equations for $\dot{\rho}$ and $\dot{\theta}$:
\begin{eqnarray}
\dot{\rho}= f_1(\rho,\, \theta)= &\frac{k}{4 D} \text{csch}\left(\frac{\pi  e^{-i \theta } \rho }{D}\right)
   \text{csch}\left(\frac{\pi  e^{i \theta } \rho }{D}\right) 
   \left[\cos(\theta ) \tan \left(\frac{\pi  \rho  \sin (\theta )}{D}\right) \cosh
   ^2\left(\frac{\pi  \rho  \cos (\theta )}{D}\right) \right. \nonumber \\[2mm]
  & \left. -\sin (\theta )
   \cos ^2\left(\frac{\pi  \rho  \sin (\theta )}{D}\right) \tanh
   \left(\frac{\pi  \rho  \cos (\theta )}{D}\right)\right]  \label{equa_rhop} \\[2mm]
\dot{\theta}= f_2(\rho,\, \theta)= & -\frac{k}{4
   D \rho } \text{csch}\left(\frac{\pi  e^{-i \theta } \rho }{D}\right)
   \text{csch}\left(\frac{\pi  e^{i \theta } \rho }{D}\right) \left[\cos
   (\theta ) \cos ^2\left(\frac{\pi  \rho  \sin (\theta )}{D}\right)
   \tanh \left(\frac{\pi  \rho  \cos (\theta )}{D}\right)+\right. \nonumber\\[2mm]
   &\left. +\sin (\theta )
   \tan \left(\frac{\pi  \rho  \sin (\theta )}{D}\right) \cosh
   ^2\left(\frac{\pi  \rho  \cos (\theta )}{D}\right)\right]\label{equa_thetap}
\end{eqnarray}
From equations \eqref{equa_rhop}  and \eqref{equa_thetap},  we finally 
derive the equation for $\rho'=\text{d}\rho/\text{d} \theta$ as follows:
\begin{equation}
\rho'= \frac{\dot{\rho}}{\dot{\theta}}= \frac{\rho  \sin (\theta ) \cos ^2\left(\frac{\pi  \rho  \sin (\theta
   )}{D}\right) \tanh \left(\frac{\pi  \rho  \cos (\theta
   )}{D}\right)-\rho  \cos (\theta ) \tan \left(\frac{\pi  \rho  \sin
   (\theta )}{D}\right) \cosh ^2\left(\frac{\pi  \rho  \cos (\theta
   )}{D}\right)}{\cos (\theta ) \cos ^2\left(\frac{\pi  \rho  \sin
   (\theta )}{D}\right) \tanh \left(\frac{\pi  \rho  \cos (\theta
   )}{D}\right)+\sin (\theta ) \tan \left(\frac{\pi  \rho  \sin (\theta
   )}{D}\right) \cosh ^2\left(\frac{\pi  \rho  \cos (\theta
   )}{D}\right)}\, ,\\[3mm]
   \label{equa_rho_theta}
\end{equation}
which is well-defined in  $\mathcal{A}=\{(\rho, \theta)\in \mathbb{R}^+ \times \mathbb{R} :  0 < \rho<D/2 \,\}$ because: a) all the elementary functions are well-defined    (included the function $\tan(...)$ through the condition $0 < \rho<D/2$); b) the denominator is  positive (in the first term  $\cos (\theta )  \cdot \tanh \left(\frac{\pi  \rho  \cos (\theta)}{D}\right) \geq 0$ and in the second term $\sin (\theta ) \cdot \tan \left(\frac{\pi  \rho  \sin (\theta )}{D}\right)\geq 0$) and never zero (both terms are never zero in  $\mathcal{A}$). 
 
In order to prove that the trajectory of vortex $P_1$ is a closed curve, we need to show that the function 
$\rho(\theta)$ is a continuos and periodic function. However, the integration of equation \eqref{equa_rho_theta} is a hard task to achieve.
Therefore, we choose to prove that $\rho(\theta)$ is a continuos and periodic function without finding the exact integral  of 
\eqref{equa_rho_theta}. In order to achieve this goal, we first need to recall a result from mathematical analysis, which states:
\medskip
{\theorem $\;\;$ Given a continuos and periodic function  ${\displaystyle f:\mathbb {R} \to \mathbb {R} }$ with period $T$ such that $\displaystyle \int_0^T f(x)\, {\rm d} x=0$, then the primitive function of $f(x)$ is periodic with period $T$. \label{Theorem_1}}\\[3mm]
\noindent
Having recalled Theorem \ref{Theorem_1}, we now need to prove the following theorem:

\medskip
{\theorem $\;\;$ The primitive function $\rho(\theta)$ of $\rho'(\theta)$ (as defined  in \eqref{equa_rho_theta}) 
is $C^1(\mathbb{R})$ and periodic with period at least $2 \pi$.  \label{Theorem_2}}\\[1mm]

The proof consists in three steps:\\[2mm]
\begin{itemize}
\item[a)] $\rho(\theta)$ is $C^1(\mathbb{R})$ function;\\[2mm]
\item[b)] $\rho'(\theta)$ is a periodic function, at least of period $T=2\pi$;\\[2mm]
\item[c)] $\displaystyle \int_0^{2\pi} \rho'(\theta)\; {\rm d} \theta=0$.
\end{itemize}

\medskip
\noindent

Below the proof of each step:\\[2mm]

\begin{itemize}
\item[a)] As stated in the previous sections, the complex velocity $w(z)$ 
is an analytic function, and hence the curve describing the trajectory of the vortex point $P_1$. 
This implies that the  function $\rho(\theta)$ is $C^1(\mathbb{R})$.
Moreover, we can assert that  the denominator of $\rho'(\theta)$  is $\neq 0$, or, better, 
it is easy to show  that it is always positive for  $(\rho, \theta)\, \in \mathcal{A}$. 
Indeed, the two terms in the denominator in \eqref{equa_rho_theta} are always positive 
(both for $\sin \theta$ and  $\cos \theta$ positive, negative or null).\\[2mm]
\item[b)] $\rho'(\theta)$ is a periodic function: in fact it follows directly from \eqref{equa_rho_theta} that 
\begin{equation}
\rho'(\theta+2\pi)=\rho'(\theta).
\end{equation}\\[2mm]
\item[c)] A sufficient condition  to prove the last step is that  the function $\displaystyle  \rho'(\theta)$ is an odd function in $\mathbb{R}$. The proof follows  directly from  \eqref{equa_rho_theta} after substituting $\theta$ by $-\theta$ obtaining:
\begin{equation}
	\rho'(-\theta)=-\rho'(\theta)
\end{equation}
\end{itemize}
Finally, we  apply Theorem\,\ref{Theorem_1} to our function $\rho'(\theta)$ and the theorem is proved. 
\medskip
Theorem \ref{Theorem_2} leads hence to the conclusion that $\rho(\theta+2\pi)=\rho$ and thus 
that the trajectory of vortex point $P_1$ is a closed curve.


\end{document}